\PassOptionsToPackage{table}{xcolor}
\documentclass[sigconf, nonacm]{acmart}

\usepackage{dutchcal}
\usepackage{fontawesome}
\usepackage[linesnumbered,ruled]{algorithm2e}
\usepackage{graphicx}
\usepackage{multirow}
\usepackage{subcaption} %
\usepackage{url}
\usepackage{xargs}
\usepackage{booktabs}
\usepackage{makecell}
\usepackage[colorinlistoftodos, prependcaption]{todonotes}
\usepackage{cprotect}
\usepackage{circledsteps}
\usepackage{annotate-equations}
\usepackage{cleveref}
\usepackage{caption}
\usepackage{xparse}
\usepackage[]{mdframed}
\usepackage{enumitem}
\usepackage[most]{tcolorbox}
\tcbuselibrary{skins,breakable}
\usepackage{threeparttable}

\renewcommand{\arraystretch}{1.3} 
\definecolor{headerbg}{RGB}{220,220,220} %
\definecolor{altrow}{RGB}{245,245,245}   %

\AtBeginDocument{%
  }

\NewDocumentCommand{\code}{v}{%
\texttt{\textcolor{black}{#1}}%
}

\crefformat{section}{\S#2#1#3} %
\crefformat{subsection}{\S#2#1#3}
\crefformat{subsubsection}{\S#2#1#3}

\newcommand\encircle[1]{%
  \tikz[baseline=(X.base)] 
    \node (X) [draw, shape=circle, inner sep=0, text=black] {\strut #1};%
}

\begin{document}

\title{ATLAHS: An \underline{A}pplication-centric Network Simulator \underline{T}oo\underline{l}chain for \underline{A}I, \underline{H}PC, and Distributed \underline{S}torage}

\author{Siyuan Shen}
\authornote{Both authors contributed equally to this work.}
\email{siyuan.shen@inf.ethz.ch}
\affiliation{%
  \institution{ETH Z\"urich}
  \city{Z\"urich}
  \country{Switzerland}
}
\author{Tommaso Bonato}
\authornotemark[1]
\email{tommaso.bonato@inf.ethz.ch}
\affiliation{%
  \institution{ETH Z\"urich}
  \city{Z\"urich}
  \country{Switzerland}
}

\author{Zhiyi Hu}
\email{zhiyihu@student.ethz.ch}
\affiliation{%
  \institution{ETH Z\"urich}
  \city{Z\"urich}
  \country{Switzerland}
}

\author{Pasquale Jordan}
\email{pasquale.jordan@inf.ethz.ch}
\affiliation{%
  \institution{ETH Z\"urich}
  \city{Z\"urich}
  \country{Switzerland}
}

\author{Tiancheng Chen}
\email{tiancheng.chen@inf.ethz.ch}
\affiliation{%
  \institution{ETH Z\"urich}
  \city{Z\"urich}
  \country{Switzerland}
}

\author{Torsten Hoefler}
\email{torsten.hoefler@inf.ethz.ch}
\affiliation{%
  \institution{ETH Z\"urich}
  \city{Z\"urich}
  \country{Switzerland}
}

\renewcommand{\shortauthors}{Shen et al.}

\begin{abstract}
Network simulators play a crucial role in evaluating the performance of large-scale systems. However, existing simulators rely heavily on synthetic microbenchmarks or narrowly focus on specific domains, limiting their ability to provide comprehensive performance insights. In this work, we introduce \textbf{ATLAHS}, a flexible, extensible, and open-source toolchain designed to trace real-world applications and accurately simulate their workloads. ATLAHS leverages the GOAL format to model communication and computation patterns in AI, HPC, and distributed storage applications. It supports multiple network simulation backends and handles multi-job and multi-tenant scenarios. Through extensive validation, we demonstrate that ATLAHS achieves high accuracy in simulating realistic workloads (consistently less than 5\% error), while significantly outperforming AstraSim, the current state-of-the-art AI systems simulator, in terms of simulation runtime and trace size efficiency. We further illustrate ATLAHS's utility via detailed case studies, highlighting the impact of congestion control algorithms on the performance of distributed storage systems, as well as the influence of job-placement strategies on application runtimes.
\end{abstract}

\maketitle

\section{Introduction}

\begin{figure}[!t]
\centering
\includegraphics[width=1\linewidth]{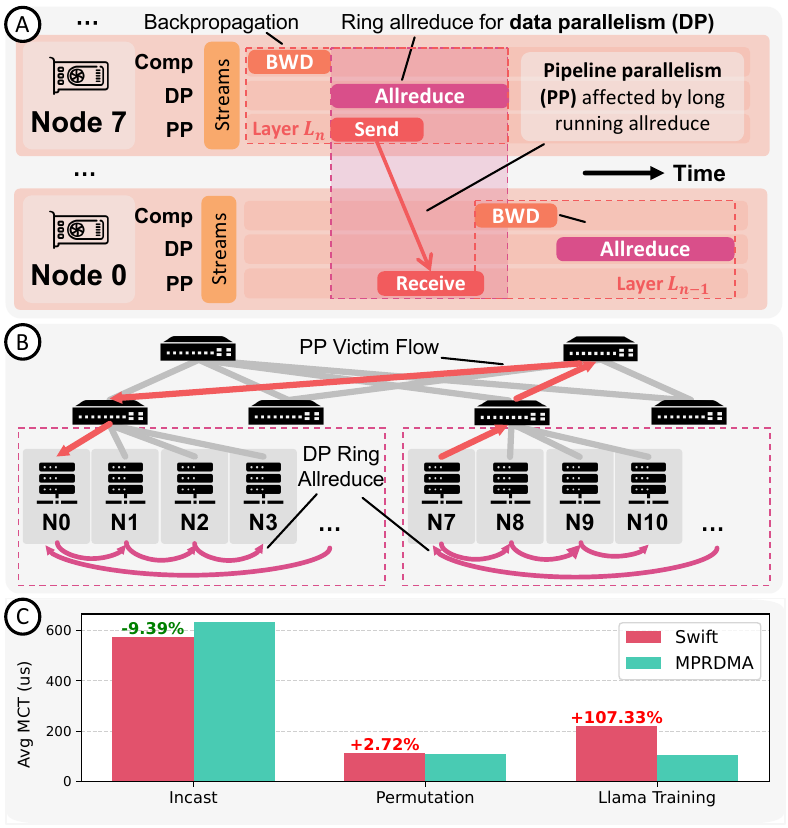}
\cprotect\caption{\protect\encircle{A} illustrates a space-time diagram of a realistic training scenario for Large Language Models (LLMs), showing overlapping communication from data parallelism (DP) and pipeline parallelism (PP). \protect\encircle{B} depicts a network-level view demonstrating how PP victim flows become congested due to simultaneous DP ring allreduce communications within a two-level fat tree topology. \protect\encircle{C} compares the performance of Swift and MPRDMA congestion control algorithms using two synthetic microbenchmarks and the LLM training workload. Percentages indicate the performance improvement (green) or degradation (red) of Swift relative to MPRDMA. }
\label{fig:poster-child}
\end{figure}

Network simulators play a critical role in evaluating the performance and feasibility of large-scale supercomputing clusters and data centers, such as Meta's \$800 million data center~\cite{meta_ai_center_wagner}, the Alps cluster at the Swiss National Supercomputing Center~\cite{cscs_alps}, or the exascale supercomputer El Capitan~\cite{el_capitan}. By creating configurable, repeatable environments that emulate traffic patterns and workloads at scale, simulators allow for rapid prototyping and enable researchers and network architects to quickly identify potential bottlenecks and performance issues without building or modifying physical infrastructure. This capability is essential for designing and optimizing complex systems before deployment.

\begin{figure*}[!t]
\centering
\includegraphics[width=1\linewidth]{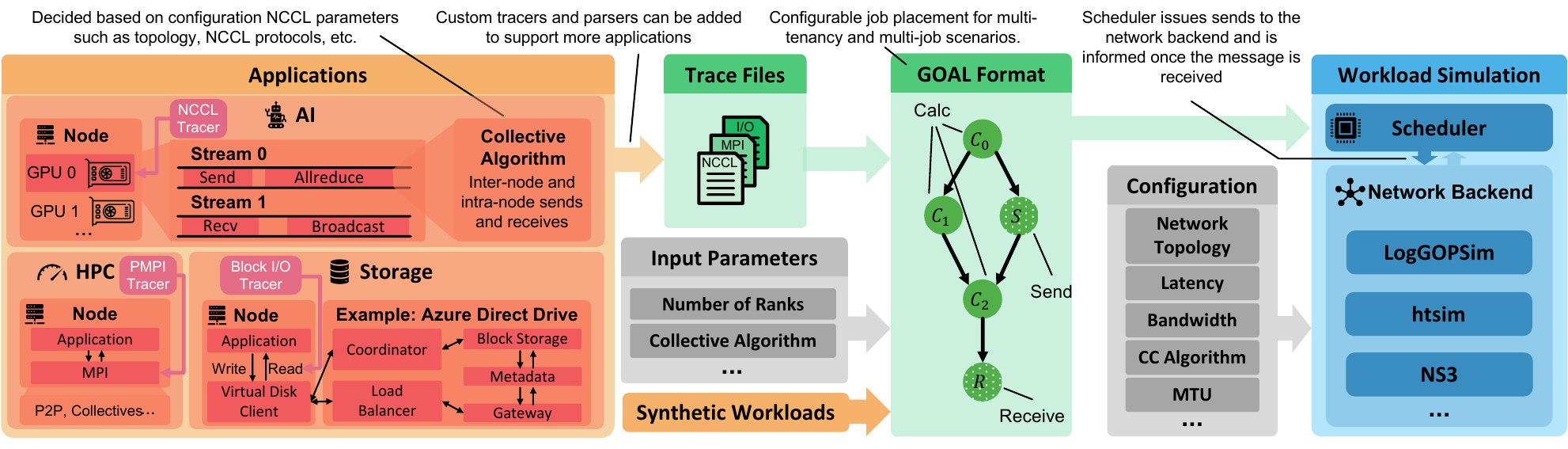}
\cprotect\caption{Overview of the ATLAHS toolchain. Application and hardware components are represented in shades of  \textcolor[HTML]{F15B5D}{red}. Trace generation and GOAL format processing are shown in \textcolor[HTML]{52B04A}{green}. Simulation is depicted in \textcolor[HTML]{4CB1E7}{blue}. For consistency, these color schemes will be used throughout the paper in all figures and diagrams.}
\label{fig:atlahs-overview}
\end{figure*}

This virtual exploration is particularly critical when developing and assessing the effectiveness of novel network topologies, protocols, and standards, such as HammingMesh~\cite{hammingmesh_hoefler}, SMaRTT-REPS~\cite{smartt_reps_bonato, reps_alone}, and Ultra Ethernet~\cite{ultra_ethernet}. For researchers and practitioners who lack access to large-scale systems, simulators provide a practical and cost-effective way to evaluate new techniques. However, many impactful networking studies primarily rely on synthetic microbenchmarks, such as incast and permutation~\cite{ndp_handley, smartt_reps_bonato, eqds_olteanu}. While useful for basic evaluations, these benchmarks often fail to accurately represent real-world workloads, potentially overlooking critical performance issues. Fig.~\ref{fig:poster-child} shows how realistic AI training workloads reveal shortcomings of the Swift~\cite{swift} congestion control algorithm, which synthetic benchmarks alone do not capture. While Swift and MPRDMA~\cite{mprdma} algorithms show comparable performance under synthetic benchmarks, AI traces analyzed with ATLAHS expose Swift’s weakness in handling multi-hop congestion due to its single end-to-end delay measurement approach. When examining the total iteration time, Swift is approximately 4\% slower, with computation partially masking the communication overhead. However, even modest slowdowns can accumulate significantly over many iterations, leading to substantial time and cost inefficiencies.

While some papers incorporate traffic generated from real microservices~\cite{flowlet_switching_vanini}, such approaches, though more realistic, often fall short of capturing the temporal dynamics, burstiness, and interdependencies inherent in real-world traffic patterns. These limitations can obscure important insights, such as correlations between traffic flows or time-varying behaviors that impact network performance.

Therefore, we emphasize that application traces are indispensable for uncovering performance issues that synthetic microbenchmarks might miss. \textit{\textbf{An application-centric approach to generating workloads for network simulators ensures that evaluations are robust and reliable in practical large-scale systems.}}

Despite this need, many state-of-the-art (SOTA) network simulators lack intuitive interfaces for parsing and replaying real application traces~\cite{ns3_riley, htsim_raiciu, cnsim_feng, codes_mubarak, ross_carothers, omnet_varga}, often requiring users to implement custom traffic generators, which significantly increases complexity and development effort. Trace-based simulators that offer built-in support for application traces tend to focus narrowly on specific domains, rather than supporting general applications. For example, AstraSim~\cite{astra-sim2_won} and SimAI~\cite{simai_wang} are tailored exclusively for AI applications, whereas LogGOPSim~\cite{loggopsim_hoefler}, PHANTOM~\cite{phantom_zhai}, and SMPI~\cite{smpi_clauss} are restricted to MPI applications in high-performance computing (HPC). Consequently, a simulation toolchain that provides a unified interface to accommodate a broad spectrum of applications would enable researchers and network engineers to conduct more thorough and versatile performance evaluations.

To this end, we introduce \textbf{\textit{ATLAHS}} (\underline{A}pplication-centric Network Simulator \underline{T}oo\underline{l}chain for \underline{A}I, \underline{H}PC, and Distributed \underline{S}torage), a toolchain designed to efficiently trace and simulate network traffic from a diverse range of applications. An overview of ATLAHS is shown in Fig.~\ref{fig:atlahs-overview}. Based on the LogGOPSim toolchain~\cite{loggopsim_hoefler}, ATLAHS leverages Group Operation Assembly Language (GOAL)~\cite{goal_hoefler}, which offers a unified representation of both computation and communication. This allows ATLAHS to not only generate synthetic microbenchmarks and traffic for applications in the aforementioned domains but also provide users with an intuitive interface to implement their own trace parsers, enabling support for any applications.

Furthermore, GOAL facilitates the integration and mixing of diverse workloads, allowing users to easily adjust workload placement to emulate multi-job and multi-tenancy scenarios. When executing simulations, ATLAHS parses GOAL files, schedules operations, and offers the flexibility to select different network simulation backends based on user requirements. Users can choose message-level simulations for faster execution or packet-level simulations for higher accuracy. We validate ATLAHS's accuracy across various backends against AstraSim, a SOTA AI simulator, and demonstrate practical use-cases through detailed case studies. These studies highlight how ATLAHS can evaluate critical factors such as the impact of congestion control algorithms on distributed storage system performance and how job-placement strategies influence application runtime. To foster open research and encourage broader adoption of ATLAHS, we publicly release a comprehensive collection of traces spanning numerous applications, domains, and configurations.

The primary contributions of this work are as follows:
\begin{enumerate}
    \item We develop ATLAHS, an open-source toolchain that emphasizes the use of application traces over synthetic microbenchmarks. This approach captures the complexity and dynamics of real-world workloads, leading to more comprehensive performance evaluations.
    \item Extending the capability of the popular LogGOPSim toolchain, ATLAHS features several novel capabilities, including the integration of AI, HPC, storage workloads, flexible simulation backends, as well as the support for multi-job and multi-tenancy scenarios.
    \item To support future research and reproducibility, we release a comprehensive collection of application traces spanning diverse domains and configurations, making them publicly available to the community.
    \item Through extensive experimentation, we validate the accuracy of the ATLAHS toolchain for a broad range of AI and HPC workloads, demonstrating that it achieves consistently high accuracy while significantly outperforming AstraSim.
    \item We present detailed case studies illustrating the versatility of ATLAHS: from analyzing the impact of congestion control algorithms on distributed storage systems to evaluating the effects of different job placement strategies on application performance within computing clusters.
\end{enumerate}

\section{Background and Related Work}

\subsection{Execution Trace Format}
\label{sec:formats}

Numerous execution trace formats exist in the field of HPC. The Open Trace Format (OTF)~\cite{otf_knupfer, otf2_knupfer}, for instance, is a trace format designed for efficient storage and special support of parallel I/O. It is used in popular HPC tools such as Score-P~\cite{scorep_knupfer}, Scalasca~\cite{scalasca_geimer}, Vampir~\cite{vampir_knupfer}, and TAU~\cite{tau_shende}. While OTF offers certain capabilities for replay in simulators~\cite{tracer_jain, otf_replay_knupfer}, its complexity presents significant challenges, as it comprises archives of files serving various purposes. This complexity likely contributes to its limited adoption beyond traditional MPI applications, particularly in domains such as AI. 

DUMPI is another widely used trace format that captures MPI communication events. It is used in the Structural Simulation Toolkit (SST) and other simulator toolchains, such as ROSS/CODES~\cite{ross_carothers, codes_mubarak}, to simulate and evaluate HPC system performance~\cite{sst_dumpi_github, sst_adalsteinsson}. However, like OTF, DUMPI is primarily tailored for MPI-based HPC applications, making it less adaptable to non-MPI domains.

Notably, most simulator toolchains rely on their own custom trace formats, including PHANTOM~\cite{phantom_zhai}, PSINS~\cite{psins_tikir}, BigSim~\cite{bigsim_zheng}, SMPI~\cite{smpi_clauss}, and SimGrid~\cite{simgrid_casanova, simgrid_2025_casanova}. However, these formats are tightly coupled to their respective simulators and are generally designed for HPC applications, particularly those based on MPI, further restricting their versatility across broader domains.

In AI, Chakra introduces a unified trace schema, Chakra ET, to capture computation and communication in machine learning (ML) applications~\cite{chakra_sridharan, proteus_duan, llmservingsim_cho, astra-sim2_won}. It also supports generative AI-based synthesis for creating workloads. However, Chakra is tightly coupled with ML-specific semantics, limiting its flexibility and excluding support for applications outside the ML domain, making it unsuitable for addressing the multi-domain requirement.

\begin{figure}[!htp]
\centering
\includegraphics[width=.85\linewidth]{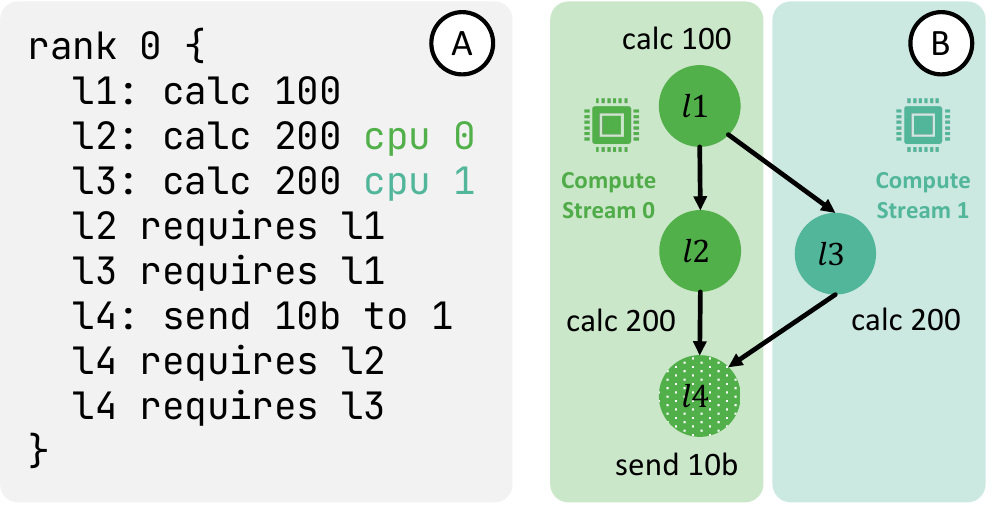}
\cprotect\caption{\protect\encircle{A} shows an example GOAL schedule of node 0 in its textual format, while \protect\encircle{B} shows the visualization of the same schedule as a DAG. Vertices in \textcolor[HTML]{52B04A}{green} are assigned to compute stream 0 to execute while vertex $l3$ in \textcolor[HTML]{51B69B}{teal} is assigned to be executed on compute stream 1.}
\label{fig:goal-example}
\end{figure}

\emph{Group Operation Assembly Language (GOAL)} was introduced as a high-level abstraction that provides a unified way to represent both computation and communication workloads in distributed and parallel systems~\cite{goal_hoefler}. GOAL defines three types of tasks: \emph{send}, \emph{receive}, and \emph{computation}. Each GOAL schedule is expressed as a directed acyclic graph (DAG), where vertices represent tasks and edges define their dependencies. Fig.~\ref{fig:goal-example} provides an example of a simple GOAL schedule.
Users can assign tasks to distinct \emph{compute streams}; if unspecified, tasks default to stream 0. This mechanism accurately represents parallel execution during simulation and facilitates the flexible distribution of workloads across multiple processing streams as specified by the user. For historical reasons, compute streams are referred to using the label \code{cpu}. In addition, to improve storage and execution efficiency, GOAL schedules are stored and executed in a compact binary format.

We chose GOAL as the intermediate trace format for ATLAHS for two main reasons. First, GOAL has been widely validated in prior work~\cite{noise_in_the_clouds_de_sensi, llamp_shen, characterizing_the_influence_hoefler, loggopsim_hoefler}, showing that its simple abstraction is sufficient to model and emulate network communication accurately. Second, its generality makes it analogous to Java bytecode, acting as a universal format to which traces from any application can be translated. This flexibility enables users to extend ATLAHS to new workloads by implementing custom tracers and parsers that produce GOAL-compliant output.

\subsection{Network Simulator Frameworks}

Over the years, numerous network simulators have been developed by industry and academia, broadly categorized into \emph{packet-level simulators}, which track individual packet traversal, and \emph{message-level simulators}, which abstract communication at the message level~\cite{towards_million_besta}. Packet-level simulators typically offer higher fidelity but incur significant computational overhead, while message-level simulators emphasize scalability and efficiency. ATLAHS, as a unified toolchain, supports both simulation approaches, allowing users to flexibly select the simulator type best suited to their requirements. In this work, we focus on widely adopted simulators: \emph{htsim}~\cite{htsim_raiciu} and \emph{NS-3}~\cite{ns3_origin, ns3_riley} as packet-level simulators, and \emph{LogGOPSim}~\cite{loggopsim_hoefler} (LGS) as the message-level simulator.

A major challenge with existing simulators is the absence of a user-friendly, general-purpose workload specification mechanism. Each simulator, sometimes even different versions of the same simulator, relies on its own workload definition approach. For instance, NS-3 typically requires workloads to be defined directly in C++, demanding in-depth knowledge of internal components and making it difficult to model complex distributed workloads. While some variants, like the HPCC-modified NS-3~\cite{hpcc}, introduce custom trace formats for datacenter workloads, they still lack intuitive support for expressing dependencies and computational overhead. These solutions are usually tailored to narrow use cases, rather than built as part of a general-purpose toolchain.

Similarly, htsim allows users to define workloads via C++ or connection matrix files in its latest version~\cite{htsim_latest}. While connection matrices provide a structured approach to workload specification, they remain limited in expressiveness, lacking support for computation modeling, an efficient tagging system for operations, and built-in trace compression.

On the other hand, LGS is a message-level simulator that abstracts communication interactions at a higher level, enabling efficient large-scale simulations. While LGS is well-suited and intended for HPC workloads, it lacks a standardized interface for broader domains, such as AI and storage systems. However, its input language, GOAL, provides all the necessary building blocks for building a generalized solution, as previously explained in Section~\ref{sec:formats}.

\section{ATLAHS Toolchain}

\subsection{Trace Collection \& GOAL Generation}

As an application-centric toolchain, ATLAHS is designed to efficiently trace applications and generate their corresponding GOAL schedules. By default, it supports tracing and GOAL generation for applications from three key domains: AI, HPC, and distributed storage, as these domains dominate the workloads in modern HPC clusters and data centers. In the following sections, we provide a detailed explanation of how traces are collected and converted into GOAL files for applications from each of these domains.

\subsubsection{\textbf{HPC}}
\label{sec:hpc-goal}

Given that ATLAHS extends LGS, which was originally designed for HPC applications, we begin with a discussion of this domain to provide context for the GOAL generation process. Notably, in the field of HPC and scientific computing, MPI remains one of the most dominant and convenient programming models~\cite{a_programming_model_performance_study_shan, exploring_traditional_karlin}. Consequently, MPI applications, as well as hybrid MPI and OpenMP applications, are the primary programming models supported for this significant category of workloads.

MPI programs are traced using a lightweight tracing library named \textbf{liballprof}, which relies on the PMPI interface to record MPI operations, their arguments, and the start and end timestamps of each operation. By analyzing the differences between timestamps of consecutive operations, the schedule generator, \textbf{Schedgen}, infers the amount of computation performed between them. Additionally, Schedgen substitutes collective MPI operations with their corresponding point-to-point (P2P) algorithms based on user specifications, enabling greater flexibility in simulation. Detailed explanations and examples of this procedure are available in~\cite{loggopsim_hoefler, llamp_shen}.

\subsubsection{\textbf{AI}}
\label{sec:ai-goal-gen}

\begin{figure}[!t]
\centering
\includegraphics[width=.9\linewidth]{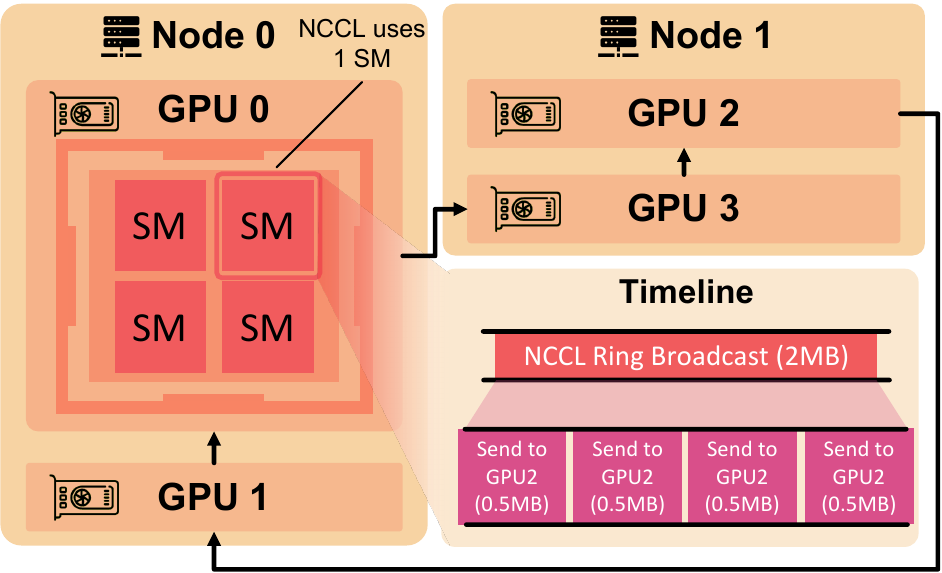}
\caption{Example of an AI application with 2 nodes and 4 GPUs connected in a ring, where each GPU communicates with its designated receiver as indicated by the arrows. NCCL utilizes a single streaming multiprocessor (SM) to handle communication operations. Using the NCCL Simple protocol, when GPU 0 broadcasts 2 MB of data as the root, the data is divided into 4 chunks, and transmitted sequentially.}
\label{fig:nccl-example}
\end{figure}

\begin{figure*}[!t]
\centering
\includegraphics[width=1\linewidth]{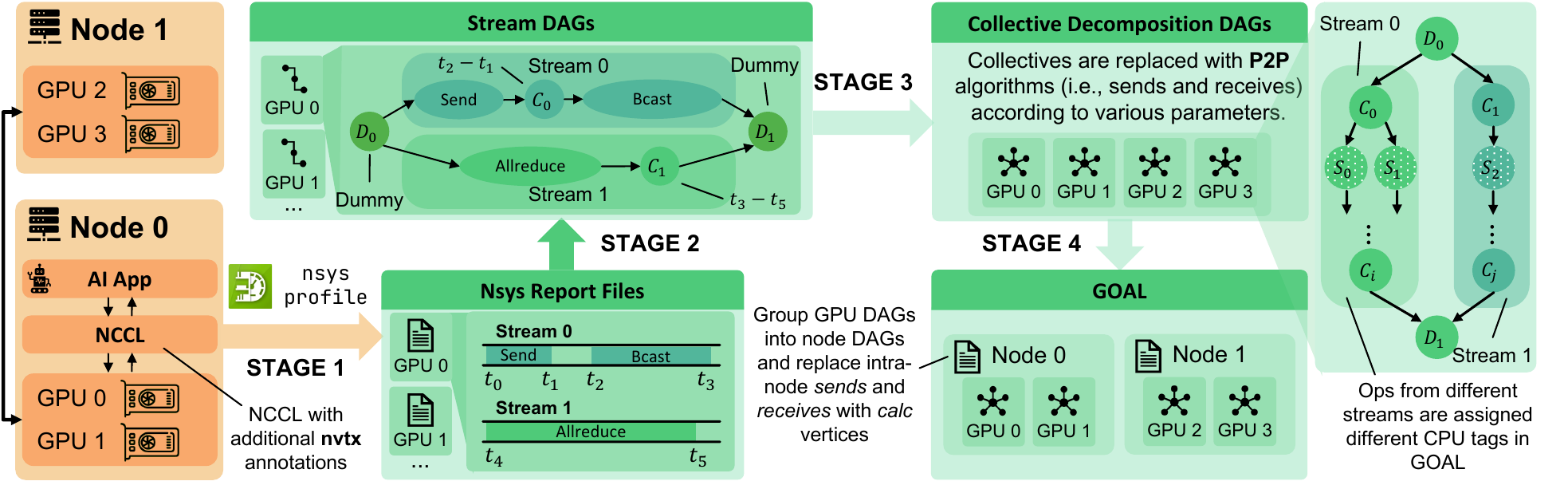}
\cprotect\caption{An example showing the 4 stages GOAL file generation for large-scale distributed AI applications.}
\label{fig:nccl-goal-gen}
\end{figure*}

To support GOAL generation for AI applications, we primarily target the NVIDIA Collective Communication Library (NCCL)~\cite{nccl_documentation} for two main reasons. First, NVIDIA's hardware and software stack accounts for over 90\% of the AI training market~\cite{nvidia_dominance_maillard, nvidia_unstoppable_ferguson}, making NCCL the de facto standard for collective communication library in most AI workloads. Second, compared to alternatives, such as like AMD's RCCL~\cite{amd_rccl_documentation} and Intel's oneCCL~\cite{intel_oneccl_documentation}, NCCL offers a more mature ecosystem of tools and profilers, which significantly accelerated our development. Note that ATLAHS is not limited to NCCL, as execution traces from other CCLs can be easily supported by implementing compatible GOAL generators.

Given the complexity of NCCL and the numerous components and configuration parameters involved, we structured the GOAL generation process into four stages, as illustrated by an example in Fig.~\ref{fig:nccl-goal-gen}. This structured approach ensures a modular design, making it easier to adapt and extend.

\paragraph{\textbf{Stage 1}}
NCCL and CUDA programming involve multiple layers and granularities of operations, with CUDA streams being the first level of parallel execution. A CUDA stream is a sequence of operations that are executed in order on the GPU. By utilizing multiple streams, developers can overlap operations, enabling concurrent execution. Therefore, the first step in our GOAL generation process is to identify the kernels executed, determine their exact execution timing, and establish dependencies and parallelism between them.

To achieve this, we use \emph{Nsight Systems}, NVIDIA’s performance analysis tool~\cite{nvidia_nsys}, to profile GPU stream activity during AI application runtime. It produces detailed nsys report files that capture operations per stream and GPU. However, key details like the communicator used by NCCL kernels are missing from the default output. To address this, we modify NCCL to add NVTX annotations~\cite{nvtx_zaitlen}, enabling collection of this information for later use. We selected Nsight Systems over custom tracers or alternatives due to its precision, efficiency, and minimal overhead, which makes it ideal for large-scale AI workloads.

\paragraph{\textbf{Stage 2}}

In the second stage, we iterate through the nsys report files for each GPU and analyze the CUDA streams. Since NCCL operations within a single stream must execute sequentially, we construct a linked list connecting each NCCL operation. Using timestamps, we then infer the computation between consecutive NCCL kernels, similar to the approach discussed in Section~\ref{sec:hpc-goal}.

To accurately represent the concurrency introduced by multiple CUDA streams, we insert dummy nodes with zero computational cost that connect the start and end vertices of each stream's operation list. Operations within different CUDA streams are assigned distinct labels, ensuring they are mapped to separate compute streams during the subsequent GOAL generation stage (details of compute streams are described in Section~\ref{sec:formats}). This explicit labeling enables the simulator to precisely model concurrent operation execution, preserving the realistic behavior of GPU-based workloads.

\paragraph{\textbf{Stage 3}}

Stage 3 is the most complex part of the GOAL generation process, as it requires decomposing NCCL collective operations into dependencies of send, receive, and computation tasks. Unlike MPI collectives, NCCL schedules vary significantly based on NCCL configuration parameters such as the number of channels, algorithm, and communication protocol, defined via \code{NCCL_MAX_NCHANNELS}, \code{NCCL_ALGO}, and \code{NCCL_PROTO}.

Fig.~\ref{fig:nccl-example} shows an example where an NCCL broadcast is decomposed into four sequential sends due to buffer size limits. If the Low Latency (LL) protocol were used instead, the schedule would differ considerably. We systematically analyzed NCCL collectives across various parameter settings and integrated the resulting schedules into ATLAHS. Due to their complexity, we omit detailed breakdowns here; full implementations are available in the source code\footnote{GitHub link: \url{https://github.com/spcl/atlahs.git}}.

\paragraph{\textbf{Stage 4}}

In the final stage, DAGs from multiple GPUs are combined to form a single DAG per node by introducing dummy nodes, following the same approach as in Stage 2. This step can be performed to reflect the original system setup, or the GPU DAGs can be restructured to explore \textbf{``what-if"} scenarios. For instance, traces from an 8-GPU, 2-node setup can be restructured to simulate a 4-node setup with 2 GPUs each, assuming the logical topology defined by \code{NCCL_ALGO} remains consistent.

Once the GPU-to-node mappings are specified, we further refine the DAG by replacing send and receive operations between GPUs on the same node with computation (calc) vertices, since intra-node communication does not traverse the inter-node fabric. The computational cost of these replacements is determined based on profiling data from the specific GPUs. For example, in Fig.~\ref{fig:nccl-goal-gen}, the communication operations between GPUs 0-1, as well as GPUs 2-3, are replaced with computation vertices.

\subsubsection{\textbf{Storage}} \label{storage_description}

\begin{figure}[!t]
\centering
\includegraphics[width=1\linewidth]{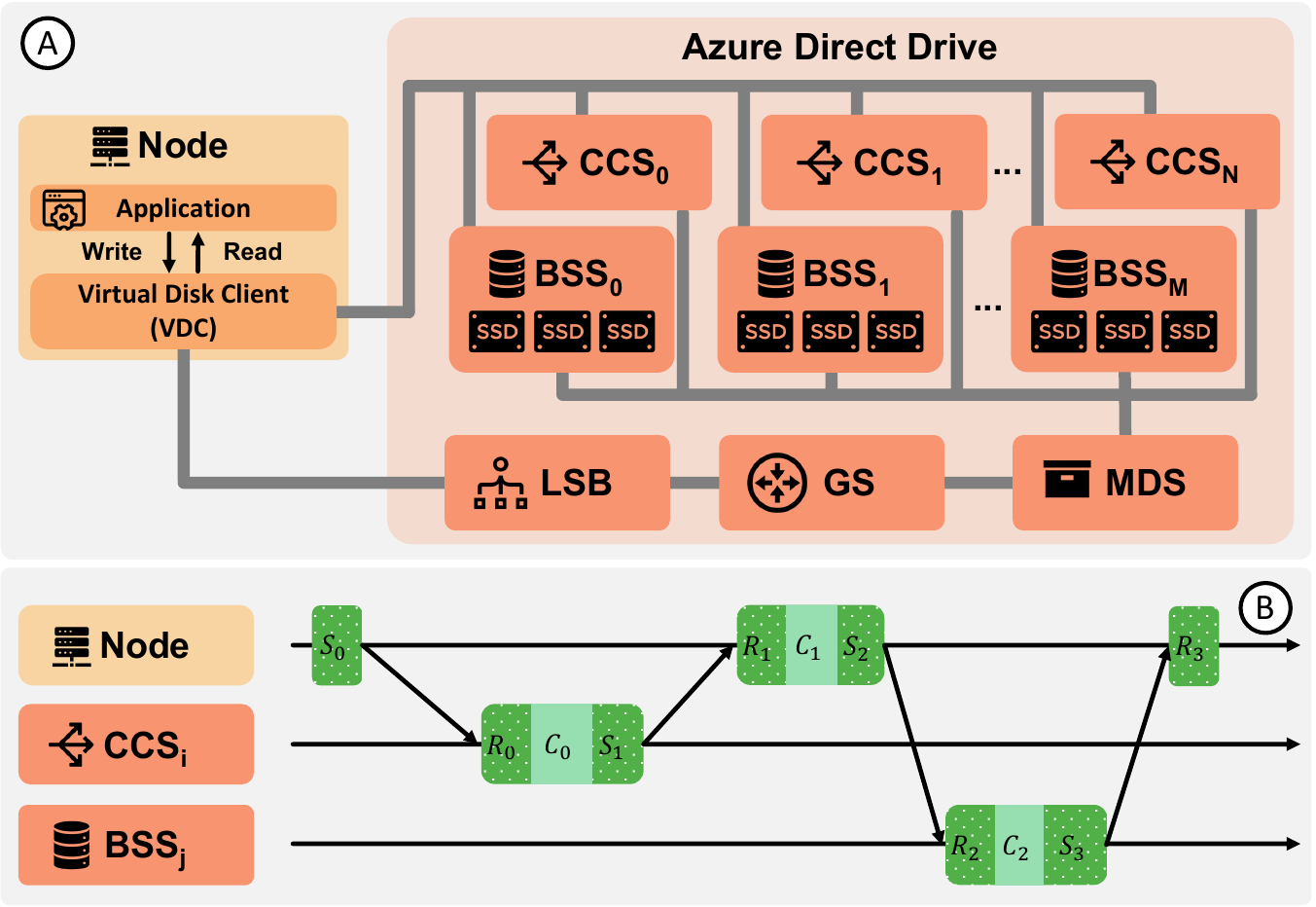}
\cprotect\caption{\protect\encircle{A} provides an overview of how applications interact with Azure Direct Drive. \protect\encircle{B} presents a space-time diagram illustrating the sequence of operations involved in a read request. Sends and receives are depicted as dotted green blocks, while computation is shown as pastel green blocks. The process begins with the node contacting the Change Coordinator Service (CCS) to determine which Block Storage Service (BSS) holds the requested data. The node then sends a request to the corresponding BSS to retrieve the data.}
\label{fig:storage-example}
\end{figure}

Distributed storage systems differ significantly from AI and HPC applications in their underlying architecture. Storage applications typically run on virtual machines (VMs) hosted by cloud providers, where disk I/O requests are issued to virtual disks backed by a distributed storage system designed for redundancy and scalability. When an application initiates a read or write request, the virtualization layer translates it into block-level operations, which the storage system processes across multiple nodes.

In this work, we focus specifically on the network communication within the storage system, capturing the underlying data transfers between nodes. For a network simulator to effectively evaluate the performance of such an architecture, it must be capable of simulating workloads and interactions between the various storage system components.

As a first step, we collect traces from arbitrary applications using a custom block-level I/O tracer built on top of bpftrace~\cite{bpftrace}, a dynamic tracing tool based on Linux’s eBPF framework~\cite{ebpf}. Unlike traditional tools like blktrace~\cite{blktrace}, which produce raw, low-level data requiring significant postprocessing, bpftrace provides a scriptable interface for filtering I/O events in real-time with minimal overhead. The resulting traces are stored in the SPC trace file format~\cite{spc_trace_file_format}, where each record corresponds to a single I/O command. This format is also used by the UMass Trace Repository~\cite{umass_storage_trace}.

I/O requests are converted into a GOAL file based on the target storage architecture. ATLAHS includes built-in support for \textbf{Azure Direct Drive}, a block storage system developed by Microsoft~\cite{direct_drive_kramer}. Fig.~\ref{fig:storage-example} provides a simplified overview, highlighting five key service components in addition to the host: Change Coordinator Service (CCS), Block Storage Service (BSS), Metadata Service (MDS), Gateway Service (GS), and Software Load Balancer (SLB). Due to space constraints, we refer readers to Microsoft's public resources for detailed descriptions~\cite{direct_drive_kramer}. As Direct Drive is proprietary, we made assumptions based on public documentation, and full implementation details are available in our open-source toolchain.

ATLAHS provides native support for Direct Drive, and its flexible and extensible framework allows network architects to evaluate a wide variety of distributed storage service architectures by implementing custom GOAL generators tailored to their own systems.

\subsection{Multi-job and Multi-tenant Scenarios} \label{support_multi_job}

To simulate multi-job workloads, where distinct applications are assigned to separate nodes and run concurrently, we simply map each application's GOAL DAG to its own nodes during GOAL generation, making this scenario easy to model. 

Multi-tenancy is common in cloud environments and is increasingly relevant in HPC and AI systems~\cite{multi_tenant_hpc_zahid, multi_tenant_ml_li}. Because GOAL represents workloads as directed acyclic graphs (DAGs), it naturally supports modeling multi-tenant workloads. By merging DAGs from different applications and introducing dummy vertices, following the approach used in Stages 2 and 4 of Section~\ref{sec:ai-goal-gen}, ATLAHS can simulate concurrency on shared nodes, enabling realistic evaluation of resource contention and communication overlap.

It is important to note that while this approach effectively models network contention in a multi-tenant environment, it does not fully capture the complexities introduced by virtualization overhead, memory subsystem interactions, or cache contention effects. Nonetheless, this method provides a lightweight and practical way to approximate multi-tenancy and analyze the resulting traffic patterns and their impact on application performance.

\subsection{Integration with Network Simulators}
\label{sec:api_interface}

\begin{figure}[!htp]
\centering
\includegraphics[width=\linewidth]{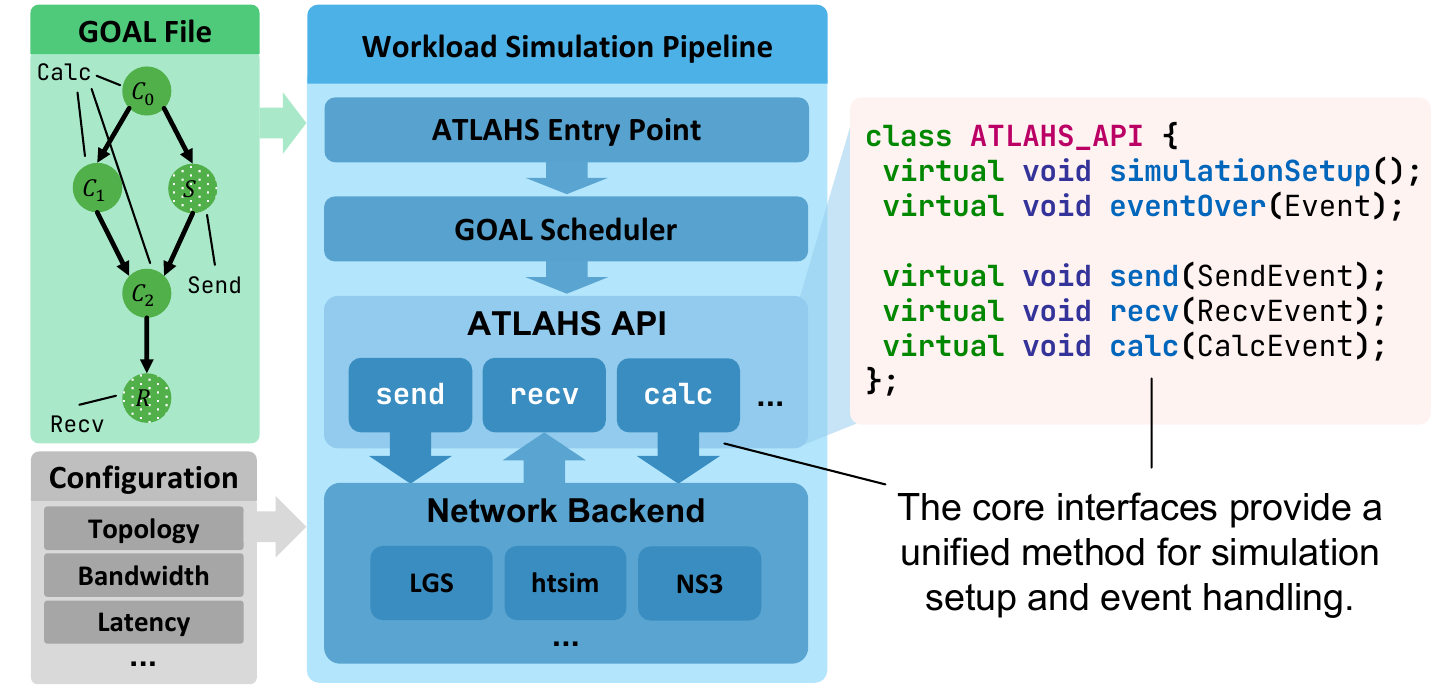}
\caption{ALTAHS APIs and an overview of its code. In the figure on the left, we only indicate the 3 core operations.}
\label{fig:atlahs_api}
\end{figure}

One of the design goals of ATLAHS is to provide a flexible toolchain that can be easily integrated with a wide range of existing network simulators. To achieve this, we abstract away simulator-specific details through a unified interface that handles a minimal set of core operations: \code{send}, \code{recv}, \code{calc}, and a helper function called \code{eventOver}, which synchronizes simulation time with ATLAHS. Each operation can be implemented to target a particular simulator backend. Additionally, a simulator-specific initialization function, \code{simulationSetup}, configures aspects such as topology, congestion control, and load balancing algorithms. Fig.~\ref{fig:atlahs_api} illustrates this integration mechanism along with its corresponding pseudocode.

Just to provide an example, the full ATLAHS interface for htsim consists of about 350 lines of code, mostly used to implement the three core operations previously defined. Depending on the specific network simulator this can be enough to cover most cases, although some simulators may require adding some corner cases when running specific scenarios.

One key aspect to make ATLAHS work with any network simulator is that it needs to be in charge of driving the actual network simulator. To do so, we implement synchronization mechanisms to match the simulation time to the internal ATLAHS time. Our approach simply uses the \code{eventOver} function to signal ATLAHS the current actual simulation time (on top of reporting the actual event that has finished). As long as a network simulator is capable of providing this information and supports the previously mentioned operation then it can easily be supported by ATLAHS.

We release the ATLAHS documentation, APIs interface, and current backend integrations publicly on GitHub.

\section{Trace Dataset}

\begin{table}[!t]
\centering
\tiny
\renewcommand{\arraystretch}{1.05}
\resizebox{\columnwidth}{!}{%
\begin{threeparttable}
\begin{tabular}{cccc}
\hline
\textbf{App} &
  \textbf{Configuration} &
  \textbf{\begin{tabular}[c]{@{}c@{}}Trace\\ (MiB)\end{tabular}} &
  \textbf{\begin{tabular}[c]{@{}c@{}}GOAL\\ (MiB)\end{tabular}} \\ \hline
\rowcolor[HTML]{EFEFEF} 
\multicolumn{1}{c|}{\cellcolor[HTML]{EFEFEF}\textbf{DLRM}}       & \multicolumn{1}{c|}{\cellcolor[HTML]{EFEFEF}4 GPUs 4 Nodes}     & 13    & 0.765  \\ \hline
\multicolumn{1}{c|}{}                                            & \multicolumn{1}{c|}{16 GPUs 4 Nodes}                            & 243   & 242   \\
\multicolumn{1}{c|}{}                                            & \multicolumn{1}{c|}{64 GPUs 16 Nodes}                           & 1566  & 2155  \\
\multicolumn{1}{c|}{\multirow{-3}{*}{\textbf{Llama 7B}}}         & \multicolumn{1}{c|}{128 GPUs 32 Nodes}                          & 1652  & 4819  \\ \hline
\rowcolor[HTML]{EFEFEF} 
\multicolumn{1}{c|}{\cellcolor[HTML]{EFEFEF}\textbf{Llama 70B}}  & \multicolumn{1}{c|}{\cellcolor[HTML]{EFEFEF}256 GPUs 64 Nodes}  & 4451  & 3561  \\ \hline
\multicolumn{1}{c|}{\textbf{\begin{tabular}[c]{@{}c@{}}MoE (Mistral) \\ 8x7B\end{tabular}}} &
  \multicolumn{1}{c|}{64 GPUs 16 Nodes} &
  1112 &
  524 \\ \hline
\rowcolor[HTML]{EFEFEF} 
\multicolumn{1}{c|}{\cellcolor[HTML]{EFEFEF}\textbf{MoE 8x13B}} &
  \multicolumn{1}{c|}{\cellcolor[HTML]{EFEFEF}128 GPUs 32 Nodes} &
  8110 &
  10054 \\ \hline
\multicolumn{1}{c|}{\textbf{MoE 8x70B}}                          & \multicolumn{1}{c|}{256 GPUs 64 Nodes}                          & 21581 & 31902 \\ \hline
\rowcolor[HTML]{EFEFEF} 
\multicolumn{1}{c|}{\cellcolor[HTML]{EFEFEF}\textbf{CloverLeaf}} & \multicolumn{1}{c|}{\cellcolor[HTML]{EFEFEF}128 Procs 8 Nodes}  & 4.1   & 5.7   \\ \hline
\multicolumn{1}{c|}{}                                            & \multicolumn{1}{c|}{128 Procs 8 Nodes}                          & 21    & 27    \\
\multicolumn{1}{c|}{}                                            & \multicolumn{1}{c|}{512 Procs 32 Nodes}                         & 132   & 171   \\
\multicolumn{1}{c|}{\multirow{-3}{*}{\textbf{HPCG}}}             & \multicolumn{1}{c|}{1024 Procs 64 Nodes}                        & 331   & 433   \\ \hline
\rowcolor[HTML]{EFEFEF} 
\multicolumn{1}{c|}{\cellcolor[HTML]{EFEFEF}}                    & \multicolumn{1}{c|}{\cellcolor[HTML]{EFEFEF}128 Procs 8 Nodes}  & 28    & 33    \\
\rowcolor[HTML]{EFEFEF} 
\multicolumn{1}{c|}{\cellcolor[HTML]{EFEFEF}}                    & \multicolumn{1}{c|}{\cellcolor[HTML]{EFEFEF}432 Procs 27 Nodes} & 137   & 166   \\
\rowcolor[HTML]{EFEFEF} 
\multicolumn{1}{c|}{\multirow{-3}{*}{\cellcolor[HTML]{EFEFEF}\textbf{LULESH}}} &
  \multicolumn{1}{c|}{\cellcolor[HTML]{EFEFEF}1024 Procs 64 Nodes} &
  351 &
  488 \\ \hline
\multicolumn{1}{c|}{}                                            & \multicolumn{1}{c|}{128 Procs 8 Nodes}                          & 3.9   & 5.6   \\
\multicolumn{1}{c|}{}                                            & \multicolumn{1}{c|}{512 Procs 32 Nodes}                         & 16    & 22    \\
\multicolumn{1}{c|}{\multirow{-3}{*}{\textbf{LAMMPS}}}           & \multicolumn{1}{c|}{1024 Procs 64 Nodes}                        & 32    & 43    \\ \hline
\rowcolor[HTML]{EFEFEF} 
\multicolumn{1}{c|}{\cellcolor[HTML]{EFEFEF}}                    & \multicolumn{1}{c|}{\cellcolor[HTML]{EFEFEF}128 Procs 8 Nodes}  & 9.6   & 13    \\
\rowcolor[HTML]{EFEFEF} 
\multicolumn{1}{c|}{\cellcolor[HTML]{EFEFEF}}                    & \multicolumn{1}{c|}{\cellcolor[HTML]{EFEFEF}512 Procs 32 Nodes} & 51    & 65    \\
\rowcolor[HTML]{EFEFEF} 
\multicolumn{1}{c|}{\multirow{-3}{*}{\cellcolor[HTML]{EFEFEF}\textbf{ICON}}} &
  \multicolumn{1}{c|}{\cellcolor[HTML]{EFEFEF}1024 Procs 64 Nodes} &
  102 &
  130 \\ \hline
\multicolumn{1}{c|}{}                                            & \multicolumn{1}{c|}{128 Procs 8 Nodes}                          & 4.6   & 9.1   \\
\multicolumn{1}{c|}{\multirow{-2}{*}{\textbf{OpenMX}}}           & \multicolumn{1}{c|}{512 Procs 32 Nodes}                         & 32    & 59    \\ \hline
\end{tabular}%
\end{threeparttable}
}
\caption{Summary of the released execution traces and corresponding GOAL files from various applications across different system configurations.}
\label{tab:traces}
\end{table}

Realistic application traces are critical for accurate network simulation and have been widely utilized in prior studies~\cite{tracer_jain, end_to_end_wolfgang, codes_mubarak, pipesim_rausch, simulation_liang, tiresias_gu, astra-sim2_won, simai_wang, llamp_shen, loggopsim_hoefler}. However, many traces remain unpublished or primarily focus on cluster-level workflows and job scheduling~\cite{pwa_feitelson, google_trace_reiss, alibaba_zhang, atlas_trace_chard, wfcommons_deelman}, lacking the granularity required for simulating individual application traffic. To bridge this gap and foster open research, we publicly release a curated collection of large-scale application traces at \url{https://spcl.inf.ethz.ch/Research/Scalable_Networking/ATLAHS/}. The collection includes both unprocessed trace files (e.g., nsys reports, MPI traces) and corresponding GOAL representations, allowing users to experiment with and convert them into other formats if needed. Table~\ref{tab:traces} summarizes available traces, and we plan to continuously expand this repository.

\section{Validation} \label{validation}

To validate the accuracy of ATLAHS, we traced numerous AI and HPC applications and compared their measured runtimes against predictions from different network backends. For AI workloads, we additionally compared ATLAHS with AstraSim 2.0~\cite{astra-sim2_won}, the current SOTA simulator for distributed ML systems. While we intended to include a comparison with SimAI~\cite{simai_wang}, its source code was not fully publicly available at the time of writing. Furthermore, due to the lack of access to Azure's Direct Drive, we showcase ATLAHS's support for distributed storage systems through a case study presented in the next section. 

We note that for htsim we used the latest available public release as starting point, but we implement several improvements to drastically improve its performance while reducing the memory usage. From our testing, the runtime of complex traces is reduced from $10\times$ to $100\times$ the after the improvements. Due to its better performance and usage by the Ultra Ethernet Consortium (UEC)~\cite{ultra_ethernet}, we focus on the ATLAHS htsim backend over NS-3 during validation. In the results, we refer to running ATLAHS with the LogGOPSim backend as \emph{ATLAHS LGS} while ATLAHS with the htsim backend as \emph{ATLAHS htsim}.

\begin{figure*}[!t]
\centering
\includegraphics[width=\linewidth]{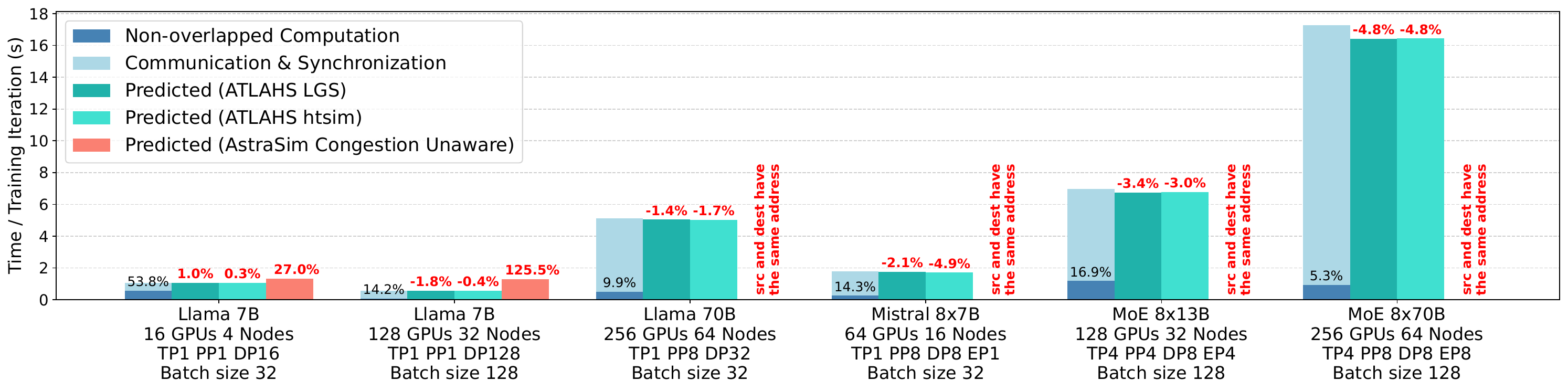}
\cprotect\caption{Comparison of measured runtimes against predicted runtimes from ATLAHS and AstraSim for various AI training workloads. The third row in the x-axis labels indicates the configuration of the parallelization strategies, where \emph{TP} stands for tensor parallelism, \emph{PP} for pipeline parallelism, \emph{DP} for data parallelism, and \emph{EP} for expert parallelism. Blue bars show actual measured runtimes, broken down into non-overlapped computation (dark blue) and communication/synchronization time (light blue). Percentages above the dark blue bars denote the proportion of non-overlapped computation in each workload, while percentages in red indicate the prediction error relative to the measured runtime.}
\label{fig:validation-ai-runtime}
\end{figure*}

\subsection{Experimental Setup}

Traces for AI workloads were collected on the Alps supercomputing cluster, operated by the Swiss National Supercomputing Center (CSCS). Alps employs a Dragonfly topology~\cite{exploring_gpu_to_gpu_desensi} and consists of 2,688 compute nodes, each featuring four NVIDIA Grace Hopper Superchips (GH200) interconnected via high-bandwidth 150 GB/s interconnect for intra-node communication and 25 GB/s per-direction Cray Slingshot interconnect for inter-node communication~\cite{understanding_gh200_fusco}. All AI workloads were executed in a containerized environment built from NVIDIA's PyTorch container (version 24.10), running on Ubuntu 22.04 with Python 3.10. We utilized a modified version of NCCL 2.20.5, extended with additional NVTX annotations.

Traces for HPC workloads were obtained from a dedicated 188-node test-bed cluster managed by CSCS. This HPC cluster is configured in a fat-tree topology using 18 Mellanox SX6036 switches. Each node is equipped with a 20-core Intel Xeon E5-2660 v2 CPU, 32 GB DDR3 RAM, and a ConnectX-3 56 Gbit/s NIC, running CentOS 7.3. The software stack utilized includes MPICH 4.1.2 and UCX 1.16.0, with the entire stack and all applications compiled using GCC 11.4.0. HPC applications were executed in a hybrid MPI+OpenMP configuration, with each node running one MPI rank complemented by 16 OpenMP threads.

Both the ATLAHS and AstraSim were executed on a dedicated machine equipped with an AMD EPYC 9654 96-core 3.7 GHz CPU and 375 GB of memory.

ATLAHS htsim employs MPRDMA~\cite{mprdma} as its congestion control mechanism, uses a buffering capacity of 1~MiB per port, and sets \(K_{\text{Min}}\) and \(K_{\text{Max}}\) to 20\% and 80\% of the queue size, respectively.

\subsection{AI}

For the AI workloads, we primarily focused on the training of Large Language Models (LLMs), as these are among the most prevalent and communication-intensive applications in modern AI. Additionally, LLM training utilizes a diverse range of parallelization strategies, making it particularly suitable for thoroughly evaluating the accuracy of ATLAHS~\cite{reasoning_llm_besta, efficient_training_llm_duan, demystifying_parallel_ben_nun}. We compared ATLAHS with AstraSim using a nightly build from February 4th, 2025. To ensure a fair evaluation, Chakra traces for AstraSim were generated directly from raw PyTorch and Kineto traces~\cite{chakra_trace_gen_rashidi}, thus guaranteeing identical execution patterns in both simulators. To reduce measurement variability, we ran each training workload for 5 warm-up iterations before collecting traces from the subsequent 2 iterations. Each experiment was conducted 5 times, and the presented results are averaged across these trials. Additionally, we calculated the percentage of non-overlapped computation to quantify the communication intensity of each workload.

Since sends and receives are executed on GPU for NCCL operations, we cannot directly obtain the LogGOPS~\cite{loggopsim_hoefler, loggp_alexandrov} parameters with tools such as Netgauge~\cite{netgauge_hoefler}, we estimated the values of the parameters from the benchmarking works of Fusco et. al~\cite{understanding_gh200_fusco}, De Sensi et al.~\cite{exploring_gpu_to_gpu_desensi}, and Groves et al.~\cite{performance_trade_offs_groves}. The final values of LogGOPS parameters are as follows: $L=3700$, $o=200$, $g=5$, $O=0$, $G=0.04$, and $S=0$, and the units will be in $ns$. The parameters we set for AstraSim emulate the real tracing setup as much as possible; details can be found in the source code we release. Throughout the experiments, we configure ATLAHS htsim to also match these parameters used by ATLAHS LGS.

Fig.~\ref{fig:validation-ai-runtime} presents validation results across various distributed training configurations using the Llama~\cite{llama_touvron} and Mixture of Experts (MoE)~\cite{moe_jiang} architectures.

Despite carefully adhering to all provided guidelines for trace generation~\cite{chakra_trace_gen_rashidi}, AstraSim only executed successfully for two configurations, encountering runtime errors in all other scenarios across different network backends. We speculate that these issues arise because AstraSim's current support for real execution traces is primarily limited to data-parallel workloads. AstraSim provides two additional backends: the congestion-aware backend and the NS-3 backend. However, the congestion-aware backend currently supports only a 1-dimensional topology, resulting in significant prediction errors when used with realistic multi-dimensional topologies, making fair comparisons infeasible. In addition, attempts to utilize the NS-3 backend consistently resulted in segmentation faults, preventing the collection of meaningful results.

We also note that ATLAHS consistently outperforms AstraSim in terms of simulation accuracy (both LGS and htsim) and speed (for LGS) for the two scenarios where AstraSim successfully executes. While not depicted in the figures, ATLAHS LGS achieves significantly shorter simulation times compared to AstraSim (Congestion Unaware backend). Specifically, in the 4-node scenario, ATLAHS LGS completes the simulation in 5.50 seconds, whereas AstraSim takes 76.63 seconds ($13.9\times$ speedup). ATLAHS htsim completes in 180.01 seconds but it is not easily comparable being a more expensive packet-level simulator. Similarly, for the 32-node scenario, ATLAHS LGS completes the simulation in 232.20 seconds compared to AstraSim's 636.87 seconds ($2.7\times$ speedup). ATLAHS htsim completes the simulation in 5100.43 seconds. All reported results represent averages across five independent trials.

\begin{figure}[!htp]
\centering
\includegraphics[width=\linewidth]{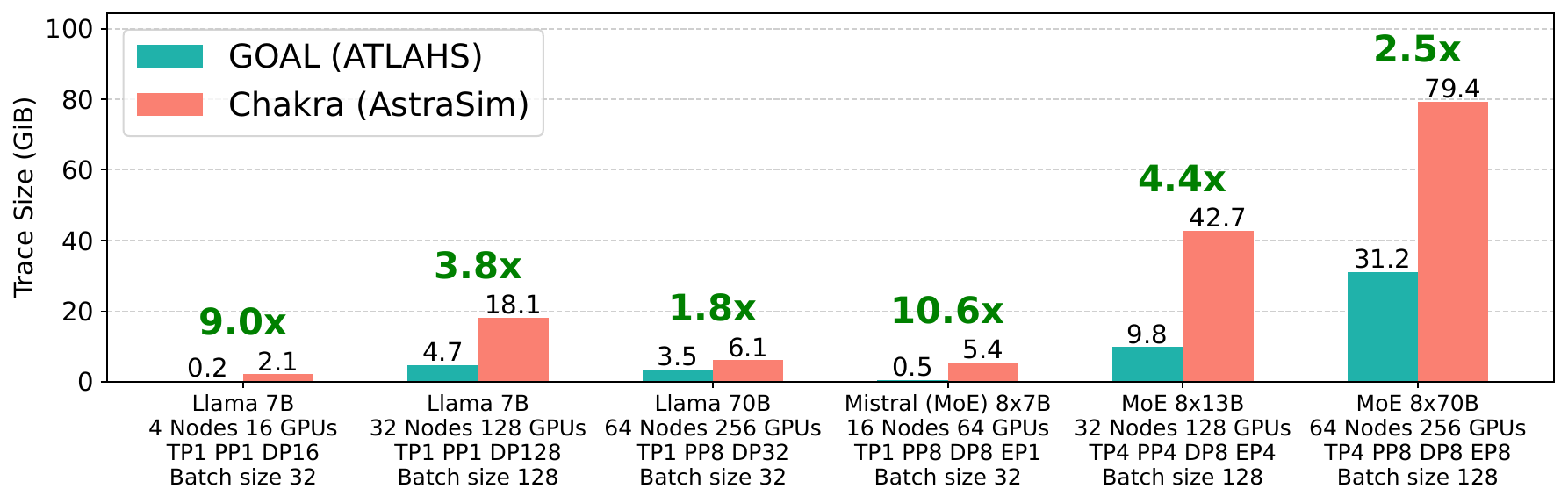}
\cprotect\caption{Trace size comparison of GOAL and Chakra. The green labels above each pair of bars indicate the trace size ratio of GOAL relative to Chakra.}
\label{fig:validation-ai-trace-size}
\end{figure}

In addition, we compared the trace sizes generated by ATLAHS and AstraSim, and the results are presented in Fig.~\ref{fig:validation-ai-trace-size}. We observe that the GOAL files utilized by ATLAHS are consistently and notably smaller than the Chakra files used by AstraSim. Although Chakra files contain additional information, such as data on compute kernels, this extra storage overhead does not appear to yield improvements in prediction accuracy.

In this section, we validated the accuracy of ATLAHS with different backends across a range of realistic LLM training scenarios in the SOTA supercomputing cluster. Furthermore, our results show that ATLAHS consistently outperforms AstraSim, one of the most popular AI system simulators, not only in terms of simulation accuracy and speed, but also in the efficiency of trace storage. These advantages highlight ATLAHS's capability as an effective toolchain for network performance evaluation for AI workloads.

\begin{figure*}[!t]
\centering
\includegraphics[width=\linewidth]{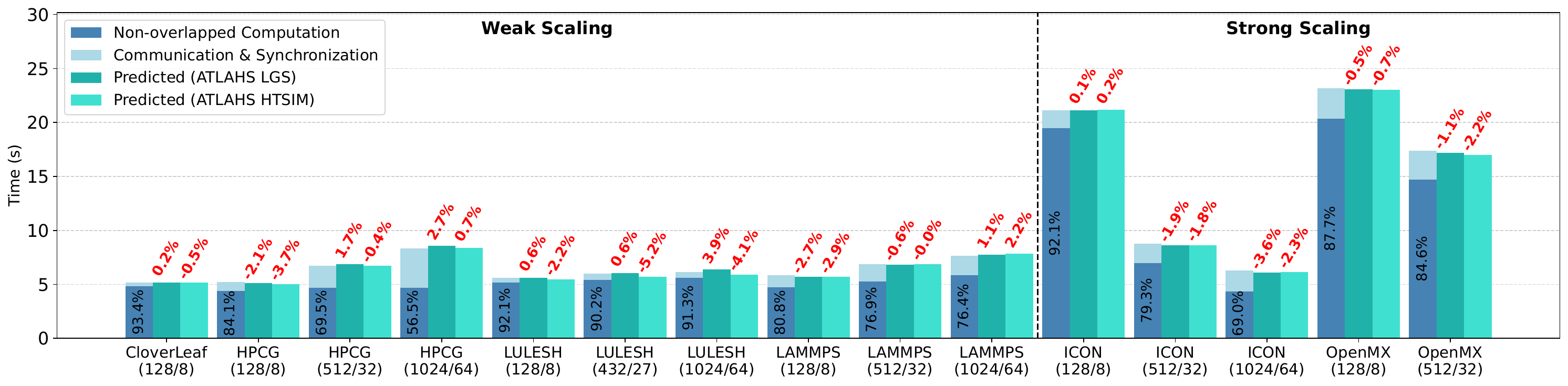}
\cprotect\caption{Comparison of measured and ATLAHS-predicted runtimes for HPC applications from various domains. The second row of the x-axis labels indicates the number of processes and nodes used in each experiment (processes/nodes). Percentages inside the dark blue bars denote the proportion of non-overlapped computation in each workload. Percentages in red above represent the prediction error of ATLAHS relative to the measured runtime.}
\label{fig:validation-hpc}
\end{figure*}

\subsection{HPC}

We measured the LogGOPS parameters using Netgauge~\cite{netgauge_hoefler}. The resulting values are $L=3000$, $o=6000$, $g=0$, $G=0.18$, $O=0$, and $S=256000$. To validate ATLAHS, we selected HPC applications spanning a wide spectrum of scientific domains, including weather and climate simulation (ICON)~\cite{icon_pham}, hydrodynamics simulation (LULESH), and molecular dynamics (LAMMPS)~\cite{lammps_thompson}, across various node configurations. For each application and configuration, the runtime was averaged over 10 independent trials, and the dataset includes both weak scaling and strong scaling scenarios.

Fig.~\ref{fig:validation-hpc} presents the validation results. It is worth noting that while the prediction error tends to increase slightly for ATLAHS LGS as the number of processes and nodes scales up, the error remains consistently below 5\% across all cases and applications. On the other hand, ATLAHS htsim does not seem to be affected negatively by the growing scale and also keeps its error rate always below 5\%. This demonstrates that ATLAHS effectively captures the underlying communication and computation dynamics across diverse HPC workloads, maintaining high accuracy over a broad range of application domains and while using different backends.

\section{Case Studies}

Experiments in this section will be designed to showcase different functionalities of ATLAHS for AI, HPC, and storage applications. Moreover, we will also focus on benefits and downsides for different ATLAHS backends.

\subsection{Effect of CC on Distributed Storage Requests} \label{storage_case_study}

\begin{figure}[!thp]
\centering
\includegraphics[width=1\linewidth]{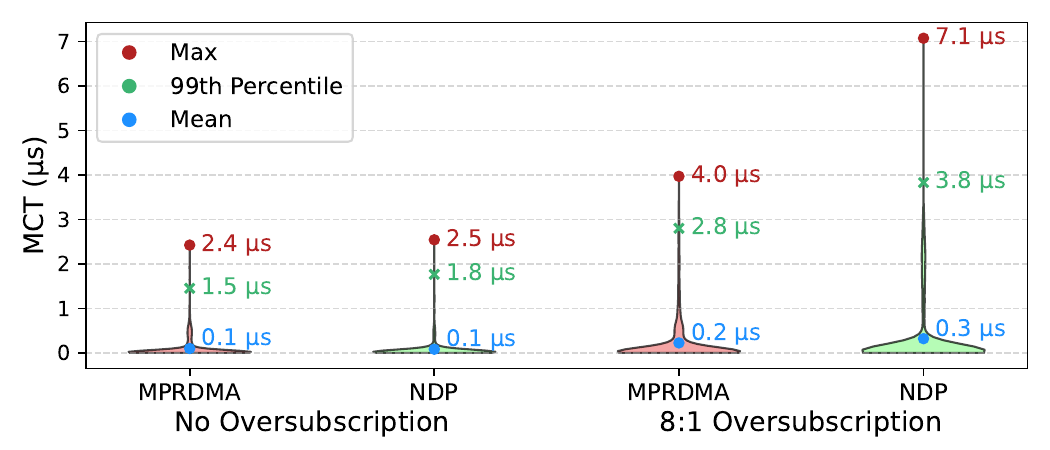}
\cprotect\caption{Comparison of the Message Completion Time (MCT) for storage traffic under different topologies and running different congestion control algorithms.}
\label{fig:case-study-storage}
\end{figure}

We now present a use case of ATLAHS related to storage traffic and the Direct Drive architecture described in Section~\ref{storage_description}. Specifically, we simulate 5k storage operations generated from the Financial distribution of the UMass collection~\cite{umass_storage_trace}. We compare two congestion control algorithms in ATLAHS htsim: MPRDMA, a sender-based algorithm, akin to DCTCP but operating on a per-packet basis, and NDP, a receiver-based algorithm. For this comparison, we employ two similar fat tree topologies: one with a fully provisioned network and one with an 8:1 oversubscription ratio between Tor switches and Core switches. In the fully provisioned topology, both algorithms perform similarly; however, in the oversubscribed case, NDP’s performance degrades significantly, with the mean Message Completion Time (MCT) increasing by 14\%, and the 99th percentile and maximum MCT rising by 35\% and 77\%, respectively. This degradation occurs because NDP, and receiver-based algorithms in general, struggle with in-network congestion occurring away from the receiver, which is evident under oversubscribed conditions. The authors of NDP acknowledge these issues, and recent work has tried to combine sender-based and receiver-based algorithms to leverage the strengths of both approaches~\cite{smartt_reps_bonato}.

This example illustrates one of many potential applications of ATLAHS for network engineers. However, since the traces that ATLAHS generates are of a general GOAL format, we envision use cases that could potentially also go behind pure networking applications or what we envisioned in this paper.

\subsection{ATLAHS LGS vs ATLAHS htsim}
In Section~\ref{validation}, we observed that the performance of ATLAHS LGS and ATLAHS htsim is generally comparable and within 1-2\% of each other for all experiments. This was only possible because of a series of assumptions that made ATLAHS LGS shine: the topology we were considering was fully provisioned and symmetric, the computation component was generally good at "masking" networking inefficiency, the collectives were designed to limit incast scenarios and we assumed no packet drops because of corruption or failures.

\begin{figure}[!htp]
\centering
\includegraphics[width=1\linewidth]{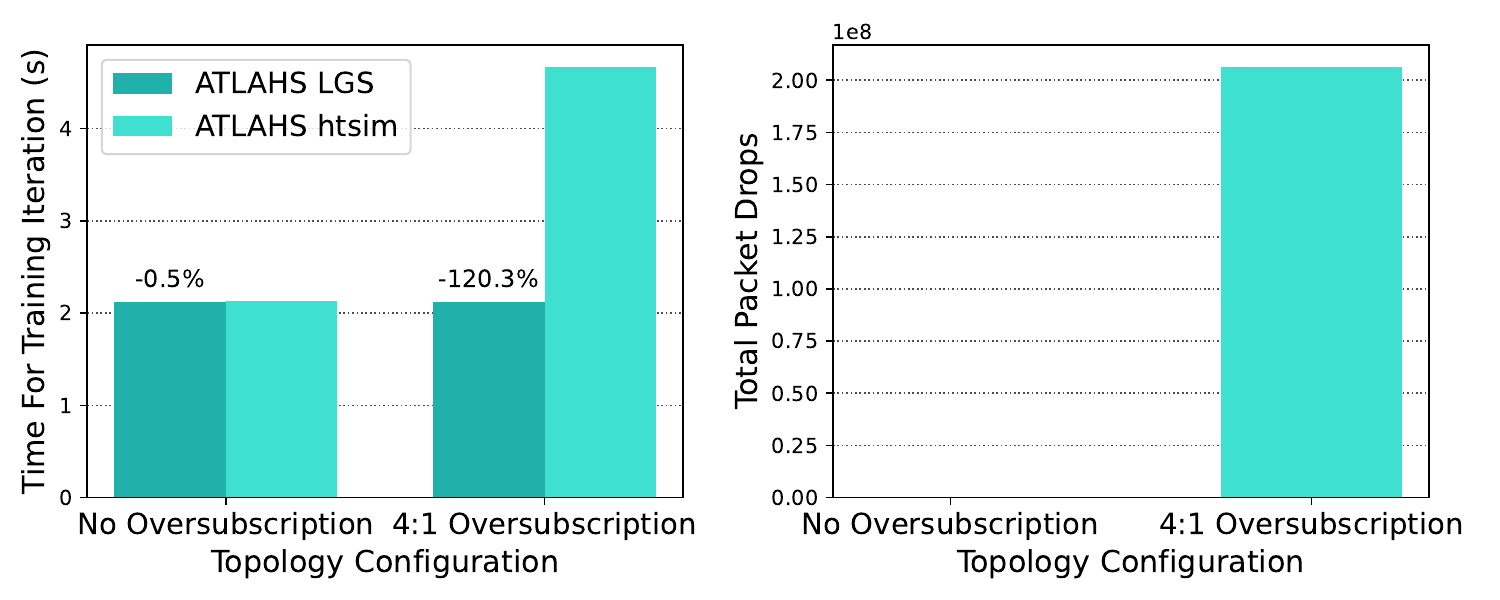}
\cprotect\caption{Comparison of the runtime of ATLAHS LGS and ATLAHS htsim when using different network topologies. The right plot showcases a possible statistic (packet losses) that only packet-level simulators can provide.}
\label{fig:lgs-vs-htsim}
\end{figure}

If these assumptions are not met, ATLAHS LGS would likely struggle, to different degrees, to provide accurate predictions.
For example, in Fig.~\ref{fig:lgs-vs-htsim}, we show this by simulating Llama 7B first on the same fully provisioned topology as before and then with a topology with a 4:1 oversubscription between the ToR and Core switches. Since ATLAHS LGS is not capable of supporting arbitrary topologies, we set $G=0.04$ for both configurations, as the theoretical injection bandwidth is unchanged even if less up-links are available in the oversubscribed topology. This naturally results in a loss of accuracy since ATLAHS LGS is oblivious to the decrease in available bandwidth from ToR to Core switches. As shown on the left of Fig.\ref{fig:lgs-vs-htsim}, we observe that, for no oversubscription, both perform well and within 1\% of each other. However, when running the 4:1 oversubscribed topology, the difference jumps to over 120\%. This is due to significant packet drops on congested uplinks (visualized on the right of Fig.~\ref{fig:lgs-vs-htsim}), which severely delay message delivery and inflate the total runtime.

Moreover, using packet-level simulators enables network engineers to gather fine-grained details, such as the total number of drops, comprehensive fairness statistics, and queue stability metrics, among other insights. Only this kind of detailed analysis, for example, enabled the analysis for storage presented in Section~\ref{storage_case_study}. However, this does not undermine the value of ATLAHS LGS. As previously demonstrated, it can deliver high accuracy under ideal conditions, and even when operating outside its ideal parameters, it provides a useful approximation with the significant benefit of being considerably faster than packet-level simulators (in most cases ATLAHS LGS is 10-50x faster than ALTAHS htsim).

\subsection{Effect of Job Placement in an HPC Cluster}
As discussed in Section~\ref{support_multi_job}, ATLAHS also provides the possibility of merging together different traces from different applications using several strategies for allocation.

\begin{figure}[!htp]
\centering
\includegraphics[width=1\linewidth]{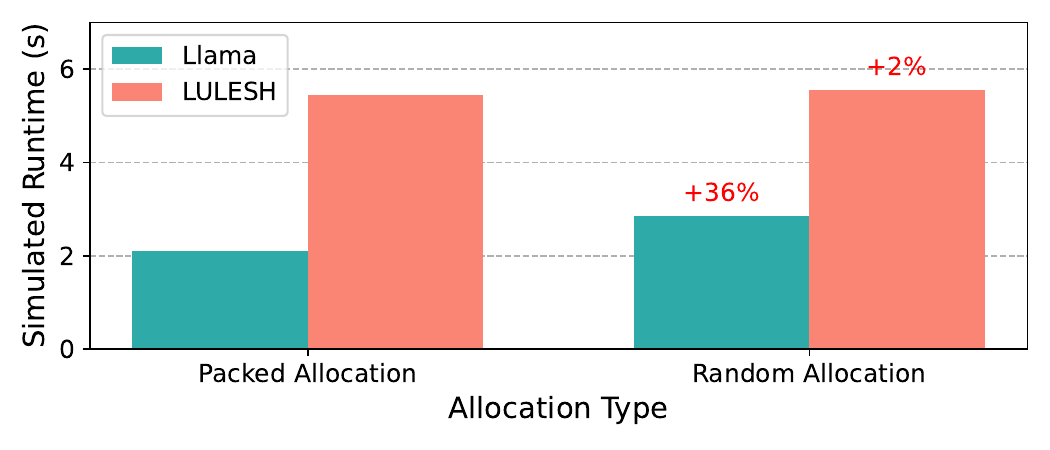}
\cprotect\caption{Comparing the runtime of different job allocation strategies when running two applications simultaneously (Llama and LULESH).}
\label{fig:case-study-job-placement}
\end{figure}

To demonstrate this capability, Fig.\ref{fig:case-study-job-placement} shows a scenario where an AI application (Llama) and an HPC application (LULESH) share a cluster. We use the same oversubscribed topology from the previous example and evaluate both workloads using ATLAHS with the htsim backend. In the “Packed Allocation” strategy, nodes are assigned sequentially to each job, keeping communication mostly local and minimizing core network usage. Conversely, in the “Random Allocation” strategy, nodes are assigned without locality, increasing inter-node distances and load on the oversubscribed core. As a result, Llama experiences a 36\% increase in runtime under random allocation. LULESH sees a smaller impact, due to its limited amount of non-overlapped computation, as shown in Fig.\ref{fig:validation-hpc}. This example highlights the value of simulating not just individual applications, but the full execution pipeline, including job placement, topology awareness, and background interference.

\section{Discussion and Extensions}

One aspect currently outside the scope of ATLAHS is detailed hardware simulation, such as modeling GPU compute kernels and interactions involving memory subsystems, which are featured in AstraSim. This exclusion is a deliberate design choice intended to prioritize the efficiency of network simulation. As demonstrated in our validation results, representing non-communication tasks simply as \code{calc} operations between communication events is sufficient to achieve accurate runtime estimations for network workloads. ATLAHS allows users to adapt traces gathered from one hardware platform to simulate another platform by applying a scaling factor derived from profiling both systems. Specifically, users can measure the relative performance difference and scale all \code{calc} values accordingly to approximate computations on different hardware.

Several areas of ATLAHS could benefit from further improvement. First, when GOAL files are generated from NCCL traces, data dependencies among CUDA kernels across streams are not currently captured. This simplification may result in inaccuracies related to computation and communication overlap, although network communication metrics remain accurate. Future work will address this by explicitly modeling CUDA kernel dependencies during GOAL generation.

Due to the nature of the GOAL format, ATLAHS currently does not support dynamically scheduled communication operations. Although our validation demonstrates that this limitation does not significantly impact the accuracy of simulating NCCL-based workloads, it may pose challenges for large-scale Graph neural networks (GNNs) training~\cite{parallel_and_distributed_gnn_besta}, programming frameworks such as \emph{Charm++}~\cite{charm_kale} or fault-tolerant protocols in distributed storage systems, where communication patterns are inherently dynamic. In future extensions, we aim to enhance GOAL by incorporating dynamic scheduling capabilities, thereby enabling ATLAHS to support a broader spectrum of applications and scenarios.

\section{Conclusion}

In this work, we introduced ATLAHS, an application-centric simulation toolchain designed to bridge the gap between realistic workload modeling and network performance evaluation across AI, HPC, and distributed storage systems. By supporting trace-based simulation through the GOAL format, ATLAHS enables accurate modeling of communication and computation patterns of a diverse spectrum of real applications. Our toolchain is highly modular and flexible, supporting multiple network simulation backends, and providing built-in support for multi-job and multi-tenant scenarios. We validated ATLAHS across a diverse set of LLM and HPC workloads, demonstrating consistently high simulation accuracy, with errors under 5\%, while outperforming SOTA frameworks such as AstraSim in both runtime efficiency and trace sizes.

Beyond validation, we demonstrated the utility of ATLAHS through detailed case studies. These highlight how congestion control algorithms can affect performance in large-scale distributed storage systems, and how job placement strategies influence performance across shared compute clusters. These insights underscore ATLAHS’s utility not only as a simulation framework, but also as a practical design and performance assessment tool for researchers and system architects seeking to optimize real-world large-scale systems under realistic workloads.

By releasing ATLAHS together with an extensive collection of application traces, we hope to foster broader community engagement and advance research into network performance evaluation. We hope ATLAHS will empower researchers and practitioners to conduct more accurate, realistic simulations, ultimately guiding the networking design of more efficient large-scale systems.

\section{Acknowledgments}

This project received funding from the European Research Council (ERC) under the European Union’s Horizon 2020 research and innovation program (grant agreement PSAP, No. 101002047). We also thank the Swiss National Supercomputing Center (CSCS) for providing the computational resources used in this work. The authors used ChatGPT-4o and 4.5 to assist with light editing and proofreading throughout the manuscript. All content and ideas are the original work of the authors.

\bibliographystyle{ACM-Reference-Format}
\bibliography{references}


\begin{thebibliography}{89}


\ifx \showCODEN    \undefined \def \showCODEN     #1{\unskip}     \fi
\ifx \showDOI      \undefined \def \showDOI       #1{#1}\fi
\ifx \showISBNx    \undefined \def \showISBNx     #1{\unskip}     \fi
\ifx \showISBNxiii \undefined \def \showISBNxiii  #1{\unskip}     \fi
\ifx \showISSN     \undefined \def \showISSN      #1{\unskip}     \fi
\ifx \showLCCN     \undefined \def \showLCCN      #1{\unskip}     \fi
\ifx \shownote     \undefined \def \shownote      #1{#1}          \fi
\ifx \showarticletitle \undefined \def \showarticletitle #1{#1}   \fi
\ifx \showURL      \undefined \def \showURL       {\relax}        \fi
\providecommand\bibfield[2]{#2}
\providecommand\bibinfo[2]{#2}
\providecommand\natexlab[1]{#1}
\providecommand\showeprint[2][]{arXiv:#2}

\bibitem[hts({[n.\,d.]})]%
        {htsim_latest}
 \bibinfo{year}{[n.\,d.]}\natexlab{}.
\newblock \bibinfo{title}{Broadcom htsim repository}.
\newblock \bibinfo{howpublished}{https://github.com/Broadcom/csg-htsim}.
\newblock
\newblock
\shownote{Accessed: 2025-02-13}.


\bibitem[ult({[n.\,d.]})]%
        {ultra_ethernet}
 \bibinfo{year}{[n.\,d.]}\natexlab{}.
\newblock \bibinfo{title}{Ultra Ethernet Consortium}.
\newblock \bibinfo{howpublished}{https://ultraethernet.org/}.
\newblock
\newblock
\shownote{Accessed: 2025-01-13}.


\bibitem[Adalsteinsson et~al\mbox{.}(2010)]%
        {sst_adalsteinsson}
\bibfield{author}{\bibinfo{person}{Helgi Adalsteinsson}, \bibinfo{person}{Scott
  Cranford}, \bibinfo{person}{David~A. Evensky}, \bibinfo{person}{Joseph~P.
  Kenny}, \bibinfo{person}{Jackson Mayo}, \bibinfo{person}{Ali Pinar}, {and}
  \bibinfo{person}{Curtis~L. Janssen}.} \bibinfo{year}{2010}\natexlab{}.
\newblock \showarticletitle{A Simulator for Large-Scale Parallel Computer
  Architectures}.
\newblock \bibinfo{journal}{\emph{Int. J. Distrib. Syst. Technol.}}
  \bibinfo{volume}{1}, \bibinfo{number}{2} (\bibinfo{date}{April}
  \bibinfo{year}{2010}), \bibinfo{pages}{57–73}.
\newblock
\showISSN{1947-3532}


\bibitem[{Advanced Micro Devices, Inc.}(2025)]%
        {amd_rccl_documentation}
\bibfield{author}{\bibinfo{person}{{Advanced Micro Devices, Inc.}}}
  \bibinfo{year}{2025}\natexlab{}.
\newblock \bibinfo{booktitle}{\emph{ROCm Communication Collectives Library
  (RCCL) Documentation}}.
\newblock
\urldef\tempurl%
\url{https://rocm.docs.amd.com/projects/rccl/en/latest/what-is-rccl.html}
\showURL{%
\tempurl}
\newblock
\shownote{Accessed: 2025-01-28}.


\bibitem[Alexandrov et~al\mbox{.}(1995)]%
        {loggp_alexandrov}
\bibfield{author}{\bibinfo{person}{Albert Alexandrov},
  \bibinfo{person}{Mihai~F. Ionescu}, \bibinfo{person}{Klaus~E. Schauser},
  {and} \bibinfo{person}{Chris Scheiman}.} \bibinfo{year}{1995}\natexlab{}.
\newblock \showarticletitle{LogGP: incorporating long messages into the LogP
  model—one step closer towards a realistic model for parallel computation}.
  In \bibinfo{booktitle}{\emph{Proceedings of the Seventh Annual ACM Symposium
  on Parallel Algorithms and Architectures}} (Santa Barbara, California, USA)
  \emph{(\bibinfo{series}{SPAA '95})}. \bibinfo{publisher}{Association for
  Computing Machinery}, \bibinfo{address}{New York, NY, USA},
  \bibinfo{pages}{95–105}.
\newblock
\showISBNx{0897917170}
\urldef\tempurl%
\url{https://doi.org/10.1145/215399.215427}
\showDOI{\tempurl}


\bibitem[Axboe(2005)]%
        {blktrace}
\bibfield{author}{\bibinfo{person}{Jens Axboe}.}
  \bibinfo{year}{2005}\natexlab{}.
\newblock \bibinfo{title}{blktrace: A Block I/O Tracing Mechanism for Linux}.
\newblock
  \bibinfo{howpublished}{\url{https://www.kernel.org/doc/Documentation/block/blktrace.txt}}.
\newblock
\newblock
\shownote{Accessed: February 04, 2025; an open‐source tool for tracing block
  I/O events in Linux}.


\bibitem[Ben-Nun and Hoefler(2019)]%
        {demystifying_parallel_ben_nun}
\bibfield{author}{\bibinfo{person}{Tal Ben-Nun} {and} \bibinfo{person}{Torsten
  Hoefler}.} \bibinfo{year}{2019}\natexlab{}.
\newblock \showarticletitle{Demystifying Parallel and Distributed Deep
  Learning: An In-depth Concurrency Analysis}.
\newblock \bibinfo{journal}{\emph{ACM Comput. Surv.}} \bibinfo{volume}{52},
  \bibinfo{number}{4}, Article \bibinfo{articleno}{65} (\bibinfo{date}{Aug.}
  \bibinfo{year}{2019}), \bibinfo{numpages}{43}~pages.
\newblock
\showISSN{0360-0300}


\bibitem[Besta et~al\mbox{.}(2025)]%
        {reasoning_llm_besta}
\bibfield{author}{\bibinfo{person}{Maciej Besta}, \bibinfo{person}{Julia
  Barth}, \bibinfo{person}{Eric Schreiber}, \bibinfo{person}{Ales Kubicek},
  \bibinfo{person}{Afonso Catarino}, \bibinfo{person}{Robert Gerstenberger},
  \bibinfo{person}{Piotr Nyczyk}, \bibinfo{person}{Patrick Iff},
  \bibinfo{person}{Yueling Li}, \bibinfo{person}{Sam Houliston},
  \bibinfo{person}{Tomasz Sternal}, \bibinfo{person}{Marcin Copik},
  \bibinfo{person}{Grzegorz Kwaśniewski}, \bibinfo{person}{Jürgen Müller},
  \bibinfo{person}{Łukasz Flis}, \bibinfo{person}{Hannes Eberhard},
  \bibinfo{person}{Hubert Niewiadomski}, {and} \bibinfo{person}{Torsten
  Hoefler}.} \bibinfo{year}{2025}\natexlab{}.
\newblock \bibinfo{title}{Reasoning Language Models: A Blueprint}.
\newblock
\newblock
\showeprint[arxiv]{2501.11223}~[cs.AI]
\urldef\tempurl%
\url{https://arxiv.org/abs/2501.11223}
\showURL{%
\tempurl}


\bibitem[Besta and Hoefler(2023)]%
        {parallel_and_distributed_gnn_besta}
\bibfield{author}{\bibinfo{person}{Maciej Besta} {and} \bibinfo{person}{Torsten
  Hoefler}.} \bibinfo{year}{2023}\natexlab{}.
\newblock \bibinfo{title}{Parallel and Distributed Graph Neural Networks: An
  In-Depth Concurrency Analysis}.
\newblock
\newblock
\showeprint[arxiv]{2205.09702}~[cs.LG]
\urldef\tempurl%
\url{https://arxiv.org/abs/2205.09702}
\showURL{%
\tempurl}


\bibitem[Besta et~al\mbox{.}(2021)]%
        {towards_million_besta}
\bibfield{author}{\bibinfo{person}{Maciej Besta}, \bibinfo{person}{Marcel
  Schneider}, \bibinfo{person}{Salvatore~Di Girolamo}, \bibinfo{person}{Ankit
  Singla}, {and} \bibinfo{person}{Torsten Hoefler}.}
  \bibinfo{year}{2021}\natexlab{}.
\newblock \bibinfo{title}{Towards Million-Server Network Simulations on Just a
  Laptop}.
\newblock
\newblock
\showeprint[arxiv]{2105.12663}~[cs.NI]
\urldef\tempurl%
\url{https://arxiv.org/abs/2105.12663}
\showURL{%
\tempurl}


\bibitem[Bonato et~al\mbox{.}(2025)]%
        {reps_alone}
\bibfield{author}{\bibinfo{person}{Tommaso Bonato}, \bibinfo{person}{Abdul
  Kabbani}, \bibinfo{person}{Ahmad Ghalayini}, \bibinfo{person}{Michael
  Papamichael}, \bibinfo{person}{Mohammad Dohadwala}, \bibinfo{person}{Lukas
  Gianinazzi}, \bibinfo{person}{Mikhail Khalilov}, \bibinfo{person}{Elias
  Achermann}, \bibinfo{person}{Daniele~De Sensi}, {and}
  \bibinfo{person}{Torsten Hoefler}.} \bibinfo{year}{2025}\natexlab{}.
\newblock \bibinfo{title}{REPS: Recycled Entropy Packet Spraying for Adaptive
  Load Balancing and Failure Mitigation}.
\newblock
\newblock
\showeprint[arxiv]{2407.21625}~[cs.NI]
\urldef\tempurl%
\url{https://arxiv.org/abs/2407.21625}
\showURL{%
\tempurl}


\bibitem[Bonato et~al\mbox{.}(2024)]%
        {smartt_reps_bonato}
\bibfield{author}{\bibinfo{person}{Tommaso Bonato}, \bibinfo{person}{Abdul
  Kabbani}, \bibinfo{person}{Daniele~De Sensi}, \bibinfo{person}{Rong Pan},
  \bibinfo{person}{Yanfang Le}, \bibinfo{person}{Costin Raiciu},
  \bibinfo{person}{Mark Handley}, \bibinfo{person}{Timo Schneider},
  \bibinfo{person}{Nils Blach}, \bibinfo{person}{Ahmad Ghalayini},
  \bibinfo{person}{Daniel Alves}, \bibinfo{person}{Michael Papamichael},
  \bibinfo{person}{Adrian Caulfield}, {and} \bibinfo{person}{Torsten Hoefler}.}
  \bibinfo{year}{2024}\natexlab{}.
\newblock \bibinfo{title}{FASTFLOW: Flexible Adaptive Congestion Control for
  High-Performance Datacenters}.
\newblock
\newblock
\showeprint[arxiv]{2404.01630}~[cs.NI]
\urldef\tempurl%
\url{https://arxiv.org/abs/2404.01630}
\showURL{%
\tempurl}


\bibitem[Carothers et~al\mbox{.}(2000)]%
        {ross_carothers}
\bibfield{author}{\bibinfo{person}{C.D. Carothers}, \bibinfo{person}{D. Bauer},
  {and} \bibinfo{person}{S. Pearce}.} \bibinfo{year}{2000}\natexlab{}.
\newblock \showarticletitle{ROSS: a high-performance, low memory, modular time
  warp system}. In \bibinfo{booktitle}{\emph{Proceedings Fourteenth Workshop on
  Parallel and Distributed Simulation}}. \bibinfo{pages}{53--60}.
\newblock
\urldef\tempurl%
\url{https://doi.org/10.1109/PADS.2000.847144}
\showDOI{\tempurl}


\bibitem[Casanova et~al\mbox{.}(2025)]%
        {simgrid_2025_casanova}
\bibfield{author}{\bibinfo{person}{Henri Casanova}, \bibinfo{person}{Arnaud
  Giersch}, \bibinfo{person}{Arnaud Legrand}, \bibinfo{person}{Martin Quinson},
  {and} \bibinfo{person}{Fr{\'e}d{\'e}ric Suter}.}
  \bibinfo{year}{2025}\natexlab{}.
\newblock \showarticletitle{{Lowering entry barriers to developing custom
  simulators of distributed applications and platforms with SimGrid}}.
\newblock \bibinfo{journal}{\emph{Parallel Comput.}}  \bibinfo{volume}{123}
  (\bibinfo{year}{2025}), \bibinfo{pages}{103--125}.
\newblock
\showISSN{0167-8191}
\urldef\tempurl%
\url{https://doi.org/10.1016/j.parco.2025.103125}
\showDOI{\tempurl}


\bibitem[Casanova et~al\mbox{.}(2008)]%
        {simgrid_casanova}
\bibfield{author}{\bibinfo{person}{Henri Casanova}, \bibinfo{person}{Arnaud
  Legrand}, {and} \bibinfo{person}{Martin Quinson}.}
  \bibinfo{year}{2008}\natexlab{}.
\newblock \showarticletitle{SimGrid: A Generic Framework for Large-Scale
  Distributed Experiments}. In \bibinfo{booktitle}{\emph{Tenth International
  Conference on Computer Modeling and Simulation (uksim 2008)}}.
  \bibinfo{pages}{126--131}.
\newblock
\urldef\tempurl%
\url{https://doi.org/10.1109/UKSIM.2008.28}
\showDOI{\tempurl}


\bibitem[Chard et~al\mbox{.}(2018)]%
        {atlas_trace_chard}
\bibfield{author}{\bibinfo{person}{Abhishek Chard}, \bibinfo{person}{R.~G.
  Dreslinski}, \bibinfo{person}{Thomas~F. Wenisch}, \bibinfo{person}{Greg
  Ganger}, {and} \bibinfo{person}{Andrew~A. Chien}.}
  \bibinfo{year}{2018}\natexlab{}.
\newblock \showarticletitle{On the Diversity of Cluster Workloads and Its
  Impact on Research Results}. In \bibinfo{booktitle}{\emph{2018 USENIX Annual
  Technical Conference (USENIX ATC 18)}}. \bibinfo{pages}{533--546}.
\newblock
\urldef\tempurl%
\url{https://www.pdl.cmu.edu/ATLAS/}
\showURL{%
\tempurl}
\newblock
\shownote{Includes LANL Mustang, Trinity, and Two Sigma traces}.


\bibitem[Cho et~al\mbox{.}(2024)]%
        {llmservingsim_cho}
\bibfield{author}{\bibinfo{person}{Jaehong Cho}, \bibinfo{person}{Minsu Kim},
  \bibinfo{person}{Hyunmin Choi}, \bibinfo{person}{Guseul Heo}, {and}
  \bibinfo{person}{Jongse Park}.} \bibinfo{year}{2024}\natexlab{}.
\newblock \showarticletitle{LLMServingSim: A HW/SW Co-Simulation Infrastructure
  for LLM Inference Serving at Scale}. In \bibinfo{booktitle}{\emph{2024 IEEE
  International Symposium on Workload Characterization (IISWC)}}.
  \bibinfo{pages}{15--29}.
\newblock
\urldef\tempurl%
\url{https://doi.org/10.1109/IISWC63097.2024.00012}
\showDOI{\tempurl}


\bibitem[Clauss et~al\mbox{.}(2011)]%
        {smpi_clauss}
\bibfield{author}{\bibinfo{person}{Pierre-Nicolas Clauss},
  \bibinfo{person}{Mark Stillwell}, \bibinfo{person}{Stephane Genaud},
  \bibinfo{person}{Frederic Suter}, \bibinfo{person}{Henri Casanova}, {and}
  \bibinfo{person}{Martin Quinson}.} \bibinfo{year}{2011}\natexlab{}.
\newblock \showarticletitle{Single Node On-Line Simulation of MPI Applications
  with SMPI}. In \bibinfo{booktitle}{\emph{2011 IEEE International Parallel \&
  Distributed Processing Symposium}}. \bibinfo{pages}{664--675}.
\newblock
\urldef\tempurl%
\url{https://doi.org/10.1109/IPDPS.2011.69}
\showDOI{\tempurl}


\bibitem[Council(2025)]%
        {spc_trace_file_format}
\bibfield{author}{\bibinfo{person}{Storage~Performance Council}.}
  \bibinfo{year}{2025}\natexlab{}.
\newblock \bibinfo{title}{SPC Trace File Format Specification}.
\newblock
  \bibinfo{howpublished}{\url{https://skulddata.cs.umass.edu/traces/storage/SPC-Traces.pdf}}.
\newblock
\newblock
\shownote{Accessed: 2025-01-22}.


\bibitem[CSCS({[n.\,d.]})]%
        {cscs_alps}
\bibfield{author}{\bibinfo{person}{CSCS}.} \bibinfo{year}{[n.\,d.]}\natexlab{}.
\newblock \showarticletitle{New Research Infrastructure: 'Alps' Supercomputer
  Inaugurated}.
\newblock \bibinfo{journal}{\emph{Swiss National Supercomputing Center}}
  (\bibinfo{year}{[n.\,d.]}).
\newblock
\urldef\tempurl%
\url{https://www.cscs.ch/publications/news/2024/new-research-infrastructure-alps-supercomputer-inaugurated}
\showURL{%
\tempurl}


\bibitem[De~Sensi et~al\mbox{.}(2022)]%
        {noise_in_the_clouds_de_sensi}
\bibfield{author}{\bibinfo{person}{Daniele De~Sensi}, \bibinfo{person}{Tiziano
  De~Matteis}, \bibinfo{person}{Konstantin Taranov}, \bibinfo{person}{Salvatore
  Di~Girolamo}, \bibinfo{person}{Tobias Rahn}, {and} \bibinfo{person}{Torsten
  Hoefler}.} \bibinfo{year}{2022}\natexlab{}.
\newblock \showarticletitle{Noise in the Clouds: Influence of Network
  Performance Variability on Application Scalability}.
\newblock \bibinfo{journal}{\emph{Proc. ACM Meas. Anal. Comput. Syst.}}
  \bibinfo{volume}{6}, \bibinfo{number}{3}, Article \bibinfo{articleno}{49}
  (\bibinfo{date}{dec} \bibinfo{year}{2022}), \bibinfo{numpages}{27}~pages.
\newblock
\urldef\tempurl%
\url{https://doi.org/10.1145/3570609}
\showDOI{\tempurl}


\bibitem[De~Sensi et~al\mbox{.}(2024)]%
        {exploring_gpu_to_gpu_desensi}
\bibfield{author}{\bibinfo{person}{Daniele De~Sensi}, \bibinfo{person}{Lorenzo
  Pichetti}, \bibinfo{person}{Flavio Vella}, \bibinfo{person}{Tiziano
  De~Matteis}, \bibinfo{person}{Zebin Ren}, \bibinfo{person}{Luigi Fusco},
  \bibinfo{person}{Matteo Turisini}, \bibinfo{person}{Daniele Cesarini},
  \bibinfo{person}{Kurt Lust}, \bibinfo{person}{Animesh Trivedi},
  \bibinfo{person}{Duncan Roweth}, \bibinfo{person}{Filippo Spiga},
  \bibinfo{person}{Salvatore Di~Girolamo}, {and} \bibinfo{person}{Torsten
  Hoefler}.} \bibinfo{year}{2024}\natexlab{}.
\newblock \showarticletitle{Exploring GPU-to-GPU Communication: Insights into
  Supercomputer Interconnects}. In \bibinfo{booktitle}{\emph{SC24:
  International Conference for High Performance Computing, Networking, Storage
  and Analysis}}. \bibinfo{publisher}{IEEE}, \bibinfo{pages}{1–15}.
\newblock
\urldef\tempurl%
\url{https://doi.org/10.1109/sc41406.2024.00039}
\showDOI{\tempurl}


\bibitem[Deelman et~al\mbox{.}(2022)]%
        {wfcommons_deelman}
\bibfield{author}{\bibinfo{person}{Ewa Deelman},
  \bibinfo{person}{Rafael~Ferreira da Silva}, \bibinfo{person}{Gideon Juve},
  \bibinfo{person}{Mats Rynge}, \bibinfo{person}{Karan Vahi}, {and}
  \bibinfo{person}{Miron Livny}.} \bibinfo{year}{2022}\natexlab{}.
\newblock \showarticletitle{WfCommons: A Framework for Enabling Scientific
  Workflow Research and Development}.
\newblock \bibinfo{journal}{\emph{Future Generation Computer Systems}}
  \bibinfo{volume}{129} (\bibinfo{year}{2022}), \bibinfo{pages}{166--182}.
\newblock
\urldef\tempurl%
\url{https://doi.org/10.1016/j.future.2022.01.011}
\showDOI{\tempurl}


\bibitem[Denzel et~al\mbox{.}(2010)]%
        {end_to_end_wolfgang}
\bibfield{author}{\bibinfo{person}{Wolfgang~E. Denzel}, \bibinfo{person}{Jian
  Li}, \bibinfo{person}{Peter Walker}, {and} \bibinfo{person}{Yuho Jin}.}
  \bibinfo{year}{2010}\natexlab{}.
\newblock \showarticletitle{A Framework for End-to-End Simulation of
  High-performance Computing Systems}.
\newblock \bibinfo{journal}{\emph{SIMULATION}} \bibinfo{volume}{86},
  \bibinfo{number}{5-6} (\bibinfo{year}{2010}), \bibinfo{pages}{331--350}.
\newblock
\urldef\tempurl%
\url{https://doi.org/10.1177/0037549709340840}
\showDOI{\tempurl}
\showeprint{https://doi.org/10.1177/0037549709340840}


\bibitem[Duan et~al\mbox{.}(2024a)]%
        {proteus_duan}
\bibfield{author}{\bibinfo{person}{Jiangfei Duan}, \bibinfo{person}{Xiuhong
  Li}, \bibinfo{person}{Ping Xu}, \bibinfo{person}{Xingcheng Zhang},
  \bibinfo{person}{Shengen Yan}, \bibinfo{person}{Yun Liang}, {and}
  \bibinfo{person}{Dahua Lin}.} \bibinfo{year}{2024}\natexlab{a}.
\newblock \showarticletitle{Proteus: Simulating the Performance of Distributed
  DNN Training}.
\newblock \bibinfo{journal}{\emph{IEEE Transactions on Parallel and Distributed
  Systems}} \bibinfo{volume}{35}, \bibinfo{number}{10} (\bibinfo{year}{2024}),
  \bibinfo{pages}{1867--1878}.
\newblock
\urldef\tempurl%
\url{https://doi.org/10.1109/TPDS.2024.3443255}
\showDOI{\tempurl}


\bibitem[Duan et~al\mbox{.}(2024b)]%
        {efficient_training_llm_duan}
\bibfield{author}{\bibinfo{person}{Jiangfei Duan}, \bibinfo{person}{Shuo
  Zhang}, \bibinfo{person}{Zerui Wang}, \bibinfo{person}{Lijuan Jiang},
  \bibinfo{person}{Wenwen Qu}, \bibinfo{person}{Qinghao Hu},
  \bibinfo{person}{Guoteng Wang}, \bibinfo{person}{Qizhen Weng},
  \bibinfo{person}{Hang Yan}, \bibinfo{person}{Xingcheng Zhang},
  \bibinfo{person}{Xipeng Qiu}, \bibinfo{person}{Dahua Lin},
  \bibinfo{person}{Yonggang Wen}, \bibinfo{person}{Xin Jin},
  \bibinfo{person}{Tianwei Zhang}, {and} \bibinfo{person}{Peng Sun}.}
  \bibinfo{year}{2024}\natexlab{b}.
\newblock \bibinfo{title}{Efficient Training of Large Language Models on
  Distributed Infrastructures: A Survey}.
\newblock
\newblock
\showeprint[arxiv]{2407.20018}~[cs.DC]
\urldef\tempurl%
\url{https://arxiv.org/abs/2407.20018}
\showURL{%
\tempurl}


\bibitem[Feitelson({[n.\,d.]})]%
        {pwa_feitelson}
\bibfield{author}{\bibinfo{person}{Dror~G. Feitelson}.}
  \bibinfo{year}{[n.\,d.]}\natexlab{}.
\newblock \bibinfo{title}{The Parallel Workloads Archive}.
\newblock
  \bibinfo{howpublished}{\url{https://www.cs.huji.ac.il/labs/parallel/workload/}}.
\newblock
\newblock
\shownote{Accessed: 2025-04-12}.


\bibitem[Feng et~al\mbox{.}(2024)]%
        {cnsim_feng}
\bibfield{author}{\bibinfo{person}{Yinxiao Feng}, \bibinfo{person}{Yuchen Wei},
  \bibinfo{person}{Dong Xiang}, {and} \bibinfo{person}{Kaisheng Ma}.}
  \bibinfo{year}{2024}\natexlab{}.
\newblock \showarticletitle{Evaluating Chiplet-based {Large-Scale}
  Interconnection Networks via {Cycle-Accurate} {Packet-Parallel} Simulation}.
  In \bibinfo{booktitle}{\emph{2024 USENIX Annual Technical Conference (USENIX
  ATC 24)}}. \bibinfo{publisher}{USENIX Association}, \bibinfo{address}{Santa
  Clara, CA}, \bibinfo{pages}{731--747}.
\newblock
\showISBNx{978-1-939133-41-0}
\urldef\tempurl%
\url{https://www.usenix.org/conference/atc24/presentation/feng-yinxiao}
\showURL{%
\tempurl}


\bibitem[Ferguson(2025)]%
        {nvidia_unstoppable_ferguson}
\bibfield{author}{\bibinfo{person}{Mackenzie Ferguson}.}
  \bibinfo{year}{2025}\natexlab{}.
\newblock \showarticletitle{Nvidia's Unstoppable Rise: Dominating the AI Chip
  Market}.
\newblock \bibinfo{journal}{\emph{OpenTools.ai}} (\bibinfo{year}{2025}).
\newblock
\urldef\tempurl%
\url{https://opentools.ai/news/nvidias-unstoppable-rise-dominating-the-ai-chip-market}
\showURL{%
\tempurl}
\newblock
\shownote{Accessed: 2025-01-28}.


\bibitem[Fusco et~al\mbox{.}(2024)]%
        {understanding_gh200_fusco}
\bibfield{author}{\bibinfo{person}{Luigi Fusco}, \bibinfo{person}{Mikhail
  Khalilov}, \bibinfo{person}{Marcin Chrapek}, \bibinfo{person}{Giridhar
  Chukkapalli}, \bibinfo{person}{Thomas Schulthess}, {and}
  \bibinfo{person}{Torsten Hoefler}.} \bibinfo{year}{2024}\natexlab{}.
\newblock \bibinfo{title}{Understanding Data Movement in Tightly Coupled
  Heterogeneous Systems: A Case Study with the Grace Hopper Superchip}.
\newblock
\newblock
\showeprint[arxiv]{2408.11556}~[cs.DC]
\urldef\tempurl%
\url{https://arxiv.org/abs/2408.11556}
\showURL{%
\tempurl}


\bibitem[Geimer et~al\mbox{.}(2010)]%
        {scalasca_geimer}
\bibfield{author}{\bibinfo{person}{Markus Geimer}, \bibinfo{person}{Felix
  Wolf}, \bibinfo{person}{Brian J.~N. Wylie}, \bibinfo{person}{Erika
  Ábrahám}, \bibinfo{person}{Daniel Becker}, {and} \bibinfo{person}{Bernd
  Mohr}.} \bibinfo{year}{2010}\natexlab{}.
\newblock \showarticletitle{The Scalasca performance toolset architecture}.
\newblock \bibinfo{journal}{\emph{Concurrency and Computation: Practice and
  Experience}}  \bibinfo{volume}{22} (\bibinfo{date}{4} \bibinfo{year}{2010}),
  \bibinfo{pages}{702--719}.
\newblock
Issue 6.
\showISSN{1532-0626}
\urldef\tempurl%
\url{https://doi.org/10.1002/cpe.1556}
\showDOI{\tempurl}


\bibitem[Groves et~al\mbox{.}(2020)]%
        {performance_trade_offs_groves}
\bibfield{author}{\bibinfo{person}{Taylor Groves}, \bibinfo{person}{Ben Brock},
  \bibinfo{person}{Yuxin Chen}, \bibinfo{person}{Khaled~Z. Ibrahim},
  \bibinfo{person}{Lenny Oliker}, \bibinfo{person}{Nicholas~J. Wright},
  \bibinfo{person}{Samuel Williams}, {and} \bibinfo{person}{Katherine Yelick}.}
  \bibinfo{year}{2020}\natexlab{}.
\newblock \showarticletitle{Performance Trade-offs in GPU Communication: A
  Study of Host and Device-initiated Approaches}. In
  \bibinfo{booktitle}{\emph{2020 IEEE/ACM Performance Modeling, Benchmarking
  and Simulation of High Performance Computer Systems (PMBS)}}.
  \bibinfo{pages}{126--137}.
\newblock
\urldef\tempurl%
\url{https://doi.org/10.1109/PMBS51919.2020.00016}
\showDOI{\tempurl}


\bibitem[Gu et~al\mbox{.}(2019)]%
        {tiresias_gu}
\bibfield{author}{\bibinfo{person}{Juncheng Gu}, \bibinfo{person}{Mosharaf
  Chowdhury}, \bibinfo{person}{Kang~G. Shin}, \bibinfo{person}{Yibo Zhu},
  \bibinfo{person}{Myeongjae Jeon}, \bibinfo{person}{Junjie Qian},
  \bibinfo{person}{Hongqiang Liu}, {and} \bibinfo{person}{Chuanxiong Guo}.}
  \bibinfo{year}{2019}\natexlab{}.
\newblock \showarticletitle{Tiresias: A {GPU} Cluster Manager for Distributed
  Deep Learning}. In \bibinfo{booktitle}{\emph{16th USENIX Symposium on
  Networked Systems Design and Implementation (NSDI 19)}}.
  \bibinfo{publisher}{USENIX Association}, \bibinfo{address}{Boston, MA},
  \bibinfo{pages}{485--500}.
\newblock
\showISBNx{978-1-931971-49-2}
\urldef\tempurl%
\url{https://www.usenix.org/conference/nsdi19/presentation/gu}
\showURL{%
\tempurl}


\bibitem[Handley et~al\mbox{.}(2017)]%
        {ndp_handley}
\bibfield{author}{\bibinfo{person}{Mark Handley}, \bibinfo{person}{Costin
  Raiciu}, \bibinfo{person}{Alexandru Agache}, \bibinfo{person}{Andrei
  Voinescu}, \bibinfo{person}{Andrew~W. Moore}, \bibinfo{person}{Gianni
  Antichi}, {and} \bibinfo{person}{Marcin W\'{o}jcik}.}
  \bibinfo{year}{2017}\natexlab{}.
\newblock \showarticletitle{Re-architecting datacenter networks and stacks for
  low latency and high performance}. In \bibinfo{booktitle}{\emph{Proceedings
  of the Conference of the ACM Special Interest Group on Data Communication}}
  (Los Angeles, CA, USA) \emph{(\bibinfo{series}{SIGCOMM '17})}.
  \bibinfo{publisher}{Association for Computing Machinery},
  \bibinfo{address}{New York, NY, USA}, \bibinfo{pages}{29–42}.
\newblock
\showISBNx{9781450346535}
\urldef\tempurl%
\url{https://doi.org/10.1145/3098822.3098825}
\showDOI{\tempurl}


\bibitem[Henderson et~al\mbox{.}(2006)]%
        {ns3_origin}
\bibfield{author}{\bibinfo{person}{Thomas Henderson}, \bibinfo{person}{Sally
  Floyd}, {and} \bibinfo{person}{George Riley}.}
  \bibinfo{year}{2006}\natexlab{}.
\newblock \showarticletitle{ns3 Project Goals}.
\newblock \bibinfo{journal}{\emph{Workshop on NS-2: the IP Network Simulator.}}
  (\bibinfo{date}{01} \bibinfo{year}{2006}).
\newblock
\urldef\tempurl%
\url{https://doi.org/10.1145/1190455.1190468}
\showDOI{\tempurl}


\bibitem[Hoefler et~al\mbox{.}(2022)]%
        {hammingmesh_hoefler}
\bibfield{author}{\bibinfo{person}{Torsten Hoefler}, \bibinfo{person}{Tommaso
  Bonato}, \bibinfo{person}{Daniele~De Sensi}, \bibinfo{person}{Salvatore~Di
  Girolamo}, \bibinfo{person}{Shigang Li}, \bibinfo{person}{Marco Heddes},
  \bibinfo{person}{Jon Belk}, \bibinfo{person}{Deepak Goel},
  \bibinfo{person}{Miguel Castro}, {and} \bibinfo{person}{Steve Scott}.}
  \bibinfo{year}{2022}\natexlab{}.
\newblock \bibinfo{title}{HammingMesh: A Network Topology for Large-Scale Deep
  Learning}.
\newblock
\newblock
\showeprint[arxiv]{2209.01346}~[cs.DC]
\urldef\tempurl%
\url{https://arxiv.org/abs/2209.01346}
\showURL{%
\tempurl}


\bibitem[Hoefler et~al\mbox{.}(2007)]%
        {netgauge_hoefler}
\bibfield{author}{\bibinfo{person}{Torsten Hoefler}, \bibinfo{person}{Torsten
  Mehlan}, \bibinfo{person}{Andrew Lumsdaine}, {and} \bibinfo{person}{Wolfgang
  Rehm}.} \bibinfo{year}{2007}\natexlab{}.
\newblock \showarticletitle{{Netgauge: A Network Performance Measurement
  Framework}}. In \bibinfo{booktitle}{\emph{Proceedings of High Performance
  Computing and Communications, HPCC'07}} (Houston, USA),
  Vol.~\bibinfo{volume}{4782}. \bibinfo{publisher}{Springer},
  \bibinfo{pages}{659--671}.
\newblock
\showISBNx{978-3-540-75443-5}


\bibitem[Hoefler et~al\mbox{.}(2010a)]%
        {characterizing_the_influence_hoefler}
\bibfield{author}{\bibinfo{person}{Torsten Hoefler}, \bibinfo{person}{Timo
  Schneider}, {and} \bibinfo{person}{Andrew Lumsdaine}.}
  \bibinfo{year}{2010}\natexlab{a}.
\newblock \showarticletitle{Characterizing the Influence of System Noise on
  Large-Scale Applications by Simulation}. In \bibinfo{booktitle}{\emph{SC '10:
  Proceedings of the 2010 ACM/IEEE International Conference for High
  Performance Computing, Networking, Storage and Analysis}}.
  \bibinfo{pages}{1--11}.
\newblock
\urldef\tempurl%
\url{https://doi.org/10.1109/SC.2010.12}
\showDOI{\tempurl}


\bibitem[Hoefler et~al\mbox{.}(2010b)]%
        {loggopsim_hoefler}
\bibfield{author}{\bibinfo{person}{Torsten Hoefler}, \bibinfo{person}{Timo
  Schneider}, {and} \bibinfo{person}{Andrew Lumsdaine}.}
  \bibinfo{year}{2010}\natexlab{b}.
\newblock \showarticletitle{LogGOPSim: simulating large-scale applications in
  the LogGOPS model}. In \bibinfo{booktitle}{\emph{Proceedings of the 19th ACM
  International Symposium on High Performance Distributed Computing}} (Chicago,
  Illinois) \emph{(\bibinfo{series}{HPDC '10})}.
  \bibinfo{publisher}{Association for Computing Machinery},
  \bibinfo{address}{New York, NY, USA}, \bibinfo{pages}{597–604}.
\newblock
\showISBNx{9781605589428}
\urldef\tempurl%
\url{https://doi.org/10.1145/1851476.1851564}
\showDOI{\tempurl}


\bibitem[Hoefler et~al\mbox{.}(2009)]%
        {goal_hoefler}
\bibfield{author}{\bibinfo{person}{Torsten Hoefler}, \bibinfo{person}{Christian
  Siebert}, {and} \bibinfo{person}{Andrew Lumsdaine}.}
  \bibinfo{year}{2009}\natexlab{}.
\newblock \showarticletitle{Group Operation Assembly Language - A Flexible Way
  to Express Collective Communication}. In \bibinfo{booktitle}{\emph{2009
  International Conference on Parallel Processing}}. \bibinfo{pages}{574--581}.
\newblock
\urldef\tempurl%
\url{https://doi.org/10.1109/ICPP.2009.70}
\showDOI{\tempurl}


\bibitem[{Intel Corporation}(2024)]%
        {intel_oneccl_documentation}
\bibfield{author}{\bibinfo{person}{{Intel Corporation}}.}
  \bibinfo{year}{2024}\natexlab{}.
\newblock \bibinfo{booktitle}{\emph{Intel® oneAPI Collective Communications
  Library (oneCCL)}}.
\newblock
\urldef\tempurl%
\url{https://www.intel.com/content/www/us/en/docs/oneapi/programming-guide/2024-1/intel-oneapi-collective-communications-library.html}
\showURL{%
\tempurl}
\newblock
\shownote{Accessed: 2025-01-28}.


\bibitem[Jain et~al\mbox{.}(2016)]%
        {tracer_jain}
\bibfield{author}{\bibinfo{person}{Nikhil Jain}, \bibinfo{person}{Abhinav
  Bhatele}, \bibinfo{person}{Sam White}, \bibinfo{person}{Todd Gamblin}, {and}
  \bibinfo{person}{Laxmikant~V. Kale}.} \bibinfo{year}{2016}\natexlab{}.
\newblock \showarticletitle{{ Evaluating HPC Networks via Simulation of
  Parallel Workloads }}. In \bibinfo{booktitle}{\emph{SC16: International
  Conference for High Performance Computing, Networking, Storage and Analysis
  (SC)}}. \bibinfo{publisher}{IEEE Computer Society}, \bibinfo{address}{Los
  Alamitos, CA, USA}, \bibinfo{pages}{154--165}.
\newblock
\showISSN{2167-4337}
\urldef\tempurl%
\url{https://doi.org/10.1109/SC.2016.13}
\showDOI{\tempurl}


\bibitem[Jiang et~al\mbox{.}(2024)]%
        {moe_jiang}
\bibfield{author}{\bibinfo{person}{Albert~Q. Jiang}, \bibinfo{person}{Alexandre
  Sablayrolles}, \bibinfo{person}{Antoine Roux}, \bibinfo{person}{Arthur
  Mensch}, \bibinfo{person}{Blanche Savary}, \bibinfo{person}{Chris Bamford},
  \bibinfo{person}{Devendra~Singh Chaplot}, \bibinfo{person}{Diego de~las
  Casas}, \bibinfo{person}{Emma~Bou Hanna}, \bibinfo{person}{Florian Bressand},
  \bibinfo{person}{Gianna Lengyel}, \bibinfo{person}{Guillaume Bour},
  \bibinfo{person}{Guillaume Lample}, \bibinfo{person}{Lélio~Renard Lavaud},
  \bibinfo{person}{Lucile Saulnier}, \bibinfo{person}{Marie-Anne Lachaux},
  \bibinfo{person}{Pierre Stock}, \bibinfo{person}{Sandeep Subramanian},
  \bibinfo{person}{Sophia Yang}, \bibinfo{person}{Szymon Antoniak},
  \bibinfo{person}{Teven~Le Scao}, \bibinfo{person}{Théophile Gervet},
  \bibinfo{person}{Thibaut Lavril}, \bibinfo{person}{Thomas Wang},
  \bibinfo{person}{Timothée Lacroix}, {and} \bibinfo{person}{William~El
  Sayed}.} \bibinfo{year}{2024}\natexlab{}.
\newblock \bibinfo{title}{Mixtral of Experts}.
\newblock
\newblock
\showeprint[arxiv]{2401.04088}~[cs.LG]
\urldef\tempurl%
\url{https://arxiv.org/abs/2401.04088}
\showURL{%
\tempurl}


\bibitem[Kale and Krishnan(1993)]%
        {charm_kale}
\bibfield{author}{\bibinfo{person}{Laxmikant~V. Kale} {and}
  \bibinfo{person}{Sanjeev Krishnan}.} \bibinfo{year}{1993}\natexlab{}.
\newblock \showarticletitle{CHARM++: a portable concurrent object oriented
  system based on C++}.
\newblock \bibinfo{journal}{\emph{SIGPLAN Not.}} \bibinfo{volume}{28},
  \bibinfo{number}{10} (\bibinfo{date}{oct} \bibinfo{year}{1993}),
  \bibinfo{pages}{91–108}.
\newblock
\showISSN{0362-1340}
\urldef\tempurl%
\url{https://doi.org/10.1145/167962.165874}
\showDOI{\tempurl}


\bibitem[Karlin et~al\mbox{.}(2013)]%
        {exploring_traditional_karlin}
\bibfield{author}{\bibinfo{person}{Ian Karlin}, \bibinfo{person}{Abhinav
  Bhatele}, \bibinfo{person}{Jeff Keasler}, \bibinfo{person}{Bradford~L.
  Chamberlain}, \bibinfo{person}{Jonathan Cohen}, \bibinfo{person}{Zachary
  Devito}, \bibinfo{person}{Riyaz Haque}, \bibinfo{person}{Dan Laney},
  \bibinfo{person}{Edward Luke}, \bibinfo{person}{Felix Wang},
  \bibinfo{person}{David Richards}, \bibinfo{person}{Martin Schulz}, {and}
  \bibinfo{person}{Charles~H. Still}.} \bibinfo{year}{2013}\natexlab{}.
\newblock \showarticletitle{Exploring Traditional and Emerging Parallel
  Programming Models Using a Proxy Application}. In
  \bibinfo{booktitle}{\emph{2013 IEEE 27th International Symposium on Parallel
  and Distributed Processing}}. \bibinfo{pages}{919--932}.
\newblock
\urldef\tempurl%
\url{https://doi.org/10.1109/IPDPS.2013.115}
\showDOI{\tempurl}


\bibitem[Kn\"{u}pfer et~al\mbox{.}(2006)]%
        {otf_knupfer}
\bibfield{author}{\bibinfo{person}{Andreas Kn\"{u}pfer}, \bibinfo{person}{Ronny
  Brendel}, \bibinfo{person}{Holger Brunst}, \bibinfo{person}{Hartmut Mix},
  {and} \bibinfo{person}{Wolfgang~E. Nagel}.} \bibinfo{year}{2006}\natexlab{}.
\newblock \showarticletitle{Introducing the open trace format (OTF)}. In
  \bibinfo{booktitle}{\emph{Proceedings of the 6th International Conference on
  Computational Science - Volume Part II}} (Reading, UK)
  \emph{(\bibinfo{series}{ICCS'06})}. \bibinfo{publisher}{Springer-Verlag},
  \bibinfo{address}{Berlin, Heidelberg}, \bibinfo{pages}{526–533}.
\newblock
\showISBNx{3540343814}
\urldef\tempurl%
\url{https://doi.org/10.1007/11758525_71}
\showDOI{\tempurl}


\bibitem[Kn{\"u}pfer et~al\mbox{.}(2008)]%
        {vampir_knupfer}
\bibfield{author}{\bibinfo{person}{Andreas Kn{\"u}pfer},
  \bibinfo{person}{Holger Brunst}, \bibinfo{person}{Jens Doleschal},
  \bibinfo{person}{Matthias Jurenz}, \bibinfo{person}{Matthias Lieber},
  \bibinfo{person}{Holger Mickler}, \bibinfo{person}{Matthias~S. M{\"u}ller},
  {and} \bibinfo{person}{Wolfgang~E. Nagel}.} \bibinfo{year}{2008}\natexlab{}.
\newblock \showarticletitle{The Vampir Performance Analysis Tool-Set}. In
  \bibinfo{booktitle}{\emph{Tools for High Performance Computing}},
  \bibfield{editor}{\bibinfo{person}{Michael Resch}, \bibinfo{person}{Rainer
  Keller}, \bibinfo{person}{Valentin Himmler}, \bibinfo{person}{Bettina
  Krammer}, {and} \bibinfo{person}{Alexander Schulz}} (Eds.).
  \bibinfo{publisher}{Springer Berlin Heidelberg}, \bibinfo{address}{Berlin,
  Heidelberg}, \bibinfo{pages}{139--155}.
\newblock
\showISBNx{978-3-540-68564-7}


\bibitem[Kn{\"u}pfer et~al\mbox{.}(2010)]%
        {otf_replay_knupfer}
\bibfield{author}{\bibinfo{person}{Andreas Kn{\"u}pfer},
  \bibinfo{person}{Markus Geimer}, \bibinfo{person}{Johannes Spazier},
  \bibinfo{person}{Joseph Schuchart}, \bibinfo{person}{Michael Wagner},
  \bibinfo{person}{Dominic Eschweiler}, {and} \bibinfo{person}{Matthias~S.
  M{\"u}ller}.} \bibinfo{year}{2010}\natexlab{}.
\newblock \showarticletitle{A generic attribute extension to OTF and its use
  for MPI replay}.
\newblock \bibinfo{journal}{\emph{Procedia Computer Science}}
  \bibinfo{volume}{1}, \bibinfo{number}{1} (\bibinfo{year}{2010}),
  \bibinfo{pages}{2115--2124}.
\newblock
\showISSN{1877-0509}
\urldef\tempurl%
\url{https://doi.org/10.1016/j.procs.2010.04.237}
\showDOI{\tempurl}
\newblock
\shownote{ICCS 2010}.


\bibitem[Kn{\"u}pfer et~al\mbox{.}(2012)]%
        {scorep_knupfer}
\bibfield{author}{\bibinfo{person}{Andreas Kn{\"u}pfer},
  \bibinfo{person}{Christian R{\"o}ssel}, \bibinfo{person}{Dieter~an Mey},
  \bibinfo{person}{Scott Biersdorff}, \bibinfo{person}{Kai Diethelm},
  \bibinfo{person}{Dominic Eschweiler}, \bibinfo{person}{Markus Geimer},
  \bibinfo{person}{Michael Gerndt}, \bibinfo{person}{Daniel Lorenz},
  \bibinfo{person}{Allen Malony}, \bibinfo{person}{Wolfgang~E. Nagel},
  \bibinfo{person}{Yury Oleynik}, \bibinfo{person}{Peter Philippen},
  \bibinfo{person}{Pavel Saviankou}, \bibinfo{person}{Dirk Schmidl},
  \bibinfo{person}{Sameer Shende}, \bibinfo{person}{Ronny Tsch{\"u}ter},
  \bibinfo{person}{Michael Wagner}, \bibinfo{person}{Bert Wesarg}, {and}
  \bibinfo{person}{Felix Wolf}.} \bibinfo{year}{2012}\natexlab{}.
\newblock \showarticletitle{Score-P: A Joint Performance Measurement Run-Time
  Infrastructure for Periscope,Scalasca, TAU, and Vampir}. In
  \bibinfo{booktitle}{\emph{Tools for High Performance Computing 2011}},
  \bibfield{editor}{\bibinfo{person}{Holger Brunst},
  \bibinfo{person}{Matthias~S. M{\"u}ller}, \bibinfo{person}{Wolfgang~E.
  Nagel}, {and} \bibinfo{person}{Michael~M. Resch}} (Eds.).
  \bibinfo{publisher}{Springer Berlin Heidelberg}, \bibinfo{address}{Berlin,
  Heidelberg}, \bibinfo{pages}{79--91}.
\newblock
\showISBNx{978-3-642-31476-6}


\bibitem[Knüpfer et~al\mbox{.}(2006)]%
        {otf2_knupfer}
\bibfield{author}{\bibinfo{person}{Andreas Knüpfer}, \bibinfo{person}{Ronny
  Brendel}, \bibinfo{person}{Holger Brunst}, \bibinfo{person}{Hartmut Mix},
  {and} \bibinfo{person}{Wolfgang~E Nagel}.} \bibinfo{year}{2006}\natexlab{}.
\newblock \bibinfo{title}{LNCS 3992 - Introducing the Open Trace Format (OTF)}.
\newblock
\newblock
\urldef\tempurl%
\url{https://doi.org/doi:10.3233/978-1-61499-041-3-481}
\showDOI{\tempurl}


\bibitem[Kramer(2023)]%
        {direct_drive_kramer}
\bibfield{author}{\bibinfo{person}{Greg Kramer}.}
  \bibinfo{year}{2023}\natexlab{}.
\newblock \bibinfo{title}{Direct Drive - Azure’s next-generation block
  storage architecture}.
\newblock
  \bibinfo{howpublished}{\url{https://storagedeveloper.org/events/agenda/session/347}}.
\newblock
\newblock
\shownote{[Online]. Accessed: 2024-02-13}.


\bibitem[Kumar et~al\mbox{.}(2020)]%
        {swift}
\bibfield{author}{\bibinfo{person}{Gautam Kumar}, \bibinfo{person}{Nandita
  Dukkipati}, \bibinfo{person}{Keon Jang}, \bibinfo{person}{Hassan Wassel},
  \bibinfo{person}{Xian Wu}, \bibinfo{person}{Behnam Montazeri},
  \bibinfo{person}{Yaogong Wang}, \bibinfo{person}{Kevin Springborn},
  \bibinfo{person}{Christopher Alfeld}, \bibinfo{person}{Mike Ryan},
  \bibinfo{person}{David~J. Wetherall}, {and} \bibinfo{person}{Amin Vahdat}.}
  \bibinfo{year}{2020}\natexlab{}.
\newblock \showarticletitle{Swift: Delay is Simple and Effective for Congestion
  Control in the Datacenter}.
\newblock
\urldef\tempurl%
\url{https://dl.acm.org/doi/pdf/10.1145/3387514.3406591}
\showURL{%
\tempurl}


\bibitem[Laboratory({[n.\,d.]})]%
        {el_capitan}
\bibfield{author}{\bibinfo{person}{Lawrence Livermore~National Laboratory}.}
  \bibinfo{year}{[n.\,d.]}\natexlab{}.
\newblock \showarticletitle{Lawrence Livermore National Laboratory’s El
  Capitan verified as world's fastest supercomputer}.
\newblock \bibinfo{journal}{\emph{LLNL}} (\bibinfo{year}{[n.\,d.]}).
\newblock
\urldef\tempurl%
\url{https://www.llnl.gov/article/52061/lawrence-livermore-national-laboratorys-el-capitan-verified-worlds-fastest-supercomputer}
\showURL{%
\tempurl}


\bibitem[Li et~al\mbox{.}(2018)]%
        {multi_tenant_ml_li}
\bibfield{author}{\bibinfo{person}{Tian Li}, \bibinfo{person}{Jie Zhong},
  \bibinfo{person}{Ji Liu}, \bibinfo{person}{Wentao Wu}, {and}
  \bibinfo{person}{Ce Zhang}.} \bibinfo{year}{2018}\natexlab{}.
\newblock \showarticletitle{Ease.ml: towards multi-tenant resource sharing for
  machine learning workloads}.
\newblock \bibinfo{journal}{\emph{Proc. VLDB Endow.}} \bibinfo{volume}{11},
  \bibinfo{number}{5} (\bibinfo{date}{Oct.} \bibinfo{year}{2018}),
  \bibinfo{pages}{607–620}.
\newblock
\showISSN{2150-8097}
\urldef\tempurl%
\url{https://doi.org/10.1145/3177732.3177737}
\showDOI{\tempurl}


\bibitem[Li et~al\mbox{.}(2019)]%
        {hpcc}
\bibfield{author}{\bibinfo{person}{Yuliang Li}, \bibinfo{person}{Rui Miao},
  \bibinfo{person}{Hongqiang~Harry Liu}, \bibinfo{person}{Yan Zhuang},
  \bibinfo{person}{Fei Feng}, \bibinfo{person}{Lingbo Tang},
  \bibinfo{person}{Zheng Cao}, \bibinfo{person}{Ming Zhang},
  \bibinfo{person}{Frank Kelly}, \bibinfo{person}{Mohammad Alizadeh}, {and}
  \bibinfo{person}{Minlan Yu}.} \bibinfo{year}{2019}\natexlab{}.
\newblock \showarticletitle{HPCC: high precision congestion control}. In
  \bibinfo{booktitle}{\emph{Proceedings of the ACM Special Interest Group on
  Data Communication}} (Beijing, China) \emph{(\bibinfo{series}{SIGCOMM '19})}.
  \bibinfo{publisher}{Association for Computing Machinery},
  \bibinfo{address}{New York, NY, USA}, \bibinfo{pages}{44–58}.
\newblock
\showISBNx{9781450359566}
\urldef\tempurl%
\url{https://doi.org/10.1145/3341302.3342085}
\showDOI{\tempurl}


\bibitem[Liang et~al\mbox{.}(2022)]%
        {simulation_liang}
\bibfield{author}{\bibinfo{person}{Ruofan Liang}, \bibinfo{person}{Bingsheng
  He}, \bibinfo{person}{Shengen Yan}, {and} \bibinfo{person}{Peng Sun}.}
  \bibinfo{year}{2022}\natexlab{}.
\newblock \bibinfo{title}{A Simulation Platform for Multi-tenant Machine
  Learning Services on Thousands of GPUs}.
\newblock
\newblock
\showeprint[arxiv]{2201.03175}~[cs.DC]
\urldef\tempurl%
\url{https://arxiv.org/abs/2201.03175}
\showURL{%
\tempurl}


\bibitem[{Linux Kernel Documentation}(2025)]%
        {ebpf}
\bibfield{author}{\bibinfo{person}{{Linux Kernel Documentation}}.}
  \bibinfo{year}{2025}\natexlab{}.
\newblock \bibinfo{title}{Extended Berkeley Packet Filter (eBPF)}.
\newblock
  \bibinfo{howpublished}{\url{https://www.kernel.org/doc/html/latest/bpf/index.html}}.
\newblock
\newblock
\shownote{Accessed: February 04, 2025. eBPF extends the classic BPF mechanism
  to run sandboxed programs in the Linux kernel for tracing, networking, and
  more.}.


\bibitem[Lu et~al\mbox{.}(2018)]%
        {mprdma}
\bibfield{author}{\bibinfo{person}{Yuanwei Lu}, \bibinfo{person}{Guo Chen},
  \bibinfo{person}{Bojie Li}, \bibinfo{person}{Kun Tan},
  \bibinfo{person}{Yongqiang Xiong}, \bibinfo{person}{Peng Cheng},
  \bibinfo{person}{Jiansong Zhang}, \bibinfo{person}{Enhong Chen}, {and}
  \bibinfo{person}{Thomas Moscibroda}.} \bibinfo{year}{2018}\natexlab{}.
\newblock \showarticletitle{{Multi-Path} Transport for {RDMA} in Datacenters}.
  In \bibinfo{booktitle}{\emph{15th USENIX Symposium on Networked Systems
  Design and Implementation (NSDI 18)}}. \bibinfo{publisher}{USENIX
  Association}, \bibinfo{address}{Renton, WA}, \bibinfo{pages}{357--371}.
\newblock
\showISBNx{978-1-939133-01-4}
\urldef\tempurl%
\url{https://www.usenix.org/conference/nsdi18/presentation/lu}
\showURL{%
\tempurl}


\bibitem[Maillard(2024)]%
        {nvidia_dominance_maillard}
\bibfield{author}{\bibinfo{person}{Dorian Maillard}.}
  \bibinfo{year}{2024}\natexlab{}.
\newblock \bibinfo{booktitle}{\emph{Nvidia's AI market dominance: Can anyone
  mount a serious challenge?}}
\newblock
\urldef\tempurl%
\url{https://www.techradar.com/pro/nvidias-ai-market-dominance-can-anyone-mount-a-serious-challenge}
\showURL{%
\tempurl}
\newblock
\shownote{Accessed: 2025-01-28}.


\bibitem[Mubarak et~al\mbox{.}(2017)]%
        {codes_mubarak}
\bibfield{author}{\bibinfo{person}{Misbah Mubarak},
  \bibinfo{person}{Christopher~D. Carothers}, \bibinfo{person}{Robert~B. Ross},
  {and} \bibinfo{person}{Philip Carns}.} \bibinfo{year}{2017}\natexlab{}.
\newblock \showarticletitle{Enabling Parallel Simulation of Large-Scale HPC
  Network Systems}.
\newblock \bibinfo{journal}{\emph{IEEE Transactions on Parallel and Distributed
  Systems}} \bibinfo{volume}{28}, \bibinfo{number}{1} (\bibinfo{year}{2017}),
  \bibinfo{pages}{87--100}.
\newblock
\urldef\tempurl%
\url{https://doi.org/10.1109/TPDS.2016.2543725}
\showDOI{\tempurl}


\bibitem[{NVIDIA Corporation}(2025a)]%
        {nccl_documentation}
\bibfield{author}{\bibinfo{person}{{NVIDIA Corporation}}.}
  \bibinfo{year}{2025}\natexlab{a}.
\newblock \bibinfo{booktitle}{\emph{NVIDIA Collective Communications Library
  (NCCL) Documentation}}.
\newblock
\urldef\tempurl%
\url{https://docs.nvidia.com/deeplearning/nccl/user-guide/docs/index.html}
\showURL{%
\tempurl}
\newblock
\shownote{Accessed: 2025-01-28}.


\bibitem[{NVIDIA Corporation}(2025b)]%
        {nvidia_nsys}
\bibfield{author}{\bibinfo{person}{{NVIDIA Corporation}}.}
  \bibinfo{year}{2025}\natexlab{b}.
\newblock \bibinfo{booktitle}{\emph{NVIDIA Nsight Systems}}.
\newblock
\urldef\tempurl%
\url{https://developer.nvidia.com/nsight-systems}
\showURL{%
\tempurl}


\bibitem[Olteanu et~al\mbox{.}(2022)]%
        {eqds_olteanu}
\bibfield{author}{\bibinfo{person}{Vladimir Olteanu}, \bibinfo{person}{Haggai
  Eran}, \bibinfo{person}{Dragos Dumitrescu}, \bibinfo{person}{Adrian Popa},
  \bibinfo{person}{Cristi Baciu}, \bibinfo{person}{Mark Silberstein},
  \bibinfo{person}{Georgios Nikolaidis}, \bibinfo{person}{Mark Handley}, {and}
  \bibinfo{person}{Costin Raiciu}.} \bibinfo{year}{2022}\natexlab{}.
\newblock \showarticletitle{An edge-queued datagram service for all datacenter
  traffic}. In \bibinfo{booktitle}{\emph{19th USENIX Symposium on Networked
  Systems Design and Implementation (NSDI 22)}}. \bibinfo{publisher}{USENIX
  Association}, \bibinfo{address}{Renton, WA}, \bibinfo{pages}{761--777}.
\newblock
\showISBNx{978-1-939133-27-4}
\urldef\tempurl%
\url{https://www.usenix.org/conference/nsdi22/presentation/olteanu}
\showURL{%
\tempurl}


\bibitem[Pham et~al\mbox{.}(2021)]%
        {icon_pham}
\bibfield{author}{\bibinfo{person}{T.~V. Pham}, \bibinfo{person}{C. Steger},
  \bibinfo{person}{B. Rockel}, \bibinfo{person}{K. Keuler}, \bibinfo{person}{I.
  Kirchner}, \bibinfo{person}{M. Mertens}, \bibinfo{person}{D. Rieger},
  \bibinfo{person}{G. Z\"angl}, {and} \bibinfo{person}{B. Fr\"uh}.}
  \bibinfo{year}{2021}\natexlab{}.
\newblock \showarticletitle{ICON in Climate Limited-area Mode (ICON release
  version 2.6.1): a new regional climate model}.
\newblock \bibinfo{journal}{\emph{Geoscientific Model Development}}
  \bibinfo{volume}{14}, \bibinfo{number}{2} (\bibinfo{year}{2021}),
  \bibinfo{pages}{985--1005}.
\newblock
\urldef\tempurl%
\url{https://doi.org/10.5194/gmd-14-985-2021}
\showDOI{\tempurl}


\bibitem[Raiciu et~al\mbox{.}(2010)]%
        {htsim_raiciu}
\bibfield{author}{\bibinfo{person}{Costin Raiciu}, \bibinfo{person}{Christopher
  Pluntke}, \bibinfo{person}{Sebastien Barre}, \bibinfo{person}{Adam
  Greenhalgh}, \bibinfo{person}{Damon Wischik}, {and} \bibinfo{person}{Mark
  Handley}.} \bibinfo{year}{2010}\natexlab{}.
\newblock \showarticletitle{Data Center Networking with Multipath TCP}. In
  \bibinfo{booktitle}{\emph{Proceedings of the 9th ACM SIGCOMM Workshop on Hot
  Topics in Networks}} (Monterey, California)
  \emph{(\bibinfo{series}{Hotnets-IX})}. \bibinfo{publisher}{Association for
  Computing Machinery}, \bibinfo{address}{New York, NY, USA}, Article
  \bibinfo{articleno}{10}, \bibinfo{numpages}{6}~pages.
\newblock
\showISBNx{9781450304092}
\urldef\tempurl%
\url{https://doi.org/10.1145/1868447.1868457}
\showDOI{\tempurl}


\bibitem[Rashidi et~al\mbox{.}(2024)]%
        {chakra_trace_gen_rashidi}
\bibfield{author}{\bibinfo{person}{Saeed Rashidi}, \bibinfo{person}{Joongun
  Park}, \bibinfo{person}{Abhilash Kolluri}, {and} \bibinfo{person}{Taekyung
  Heo}.} \bibinfo{year}{2024}\natexlab{}.
\newblock \bibinfo{title}{Chakra Execution Trace Collection – A Comprehensive
  Guide on Merging PyTorch and Kineto Traces}.
\newblock
  \bibinfo{howpublished}{https://github.com/mlcommons/chakra/wiki/Chakra-Execution-Trace-Collection-\%E2\%80\%90-A-Comprehensive-Guide-on-Merging-PyTorch-and-Kineto-Traces}.
\newblock
\newblock
\shownote{GitHub wiki page, last edited on September 24, 2024. Accessed on
  March 26, 2025.}.


\bibitem[Rausch et~al\mbox{.}(2020)]%
        {pipesim_rausch}
\bibfield{author}{\bibinfo{person}{Thomas Rausch}, \bibinfo{person}{Waldemar
  Hummer}, {and} \bibinfo{person}{Vinod Muthusamy}.}
  \bibinfo{year}{2020}\natexlab{}.
\newblock \bibinfo{title}{PipeSim: Trace-driven Simulation of Large-Scale AI
  Operations Platforms}.
\newblock
\newblock
\showeprint[arxiv]{2006.12587}~[cs.DC]
\urldef\tempurl%
\url{https://arxiv.org/abs/2006.12587}
\showURL{%
\tempurl}


\bibitem[Reiss et~al\mbox{.}(2012)]%
        {google_trace_reiss}
\bibfield{author}{\bibinfo{person}{Charles Reiss}, \bibinfo{person}{Alexey
  Tumanov}, \bibinfo{person}{Gregory~R. Ganger}, \bibinfo{person}{Randy~H.
  Katz}, {and} \bibinfo{person}{Michael~A. Kozuch}.}
  \bibinfo{year}{2012}\natexlab{}.
\newblock \showarticletitle{Heterogeneity and dynamicity of clouds at scale:
  Google trace analysis}. In \bibinfo{booktitle}{\emph{Proceedings of the Third
  ACM Symposium on Cloud Computing (SoCC)}}. \bibinfo{pages}{7:1--7:13}.
\newblock
\urldef\tempurl%
\url{https://doi.org/10.1145/2391229.2391236}
\showDOI{\tempurl}


\bibitem[Riley and Henderson(2010)]%
        {ns3_riley}
\bibfield{author}{\bibinfo{person}{George~F Riley} {and}
  \bibinfo{person}{Thomas~R Henderson}.} \bibinfo{year}{2010}\natexlab{}.
\newblock \showarticletitle{The ns-3 network simulator}.
\newblock In \bibinfo{booktitle}{\emph{Modeling and tools for network
  simulation}}. \bibinfo{publisher}{Springer}, \bibinfo{pages}{15--34}.
\newblock


\bibitem[Robertson and contributors(2025)]%
        {bpftrace}
\bibfield{author}{\bibinfo{person}{Alastair Robertson} {and}
  \bibinfo{person}{contributors}.} \bibinfo{year}{2025}\natexlab{}.
\newblock \bibinfo{title}{{bpftrace: A High-Level Tracing Language for Linux}}.
\newblock
\newblock
\urldef\tempurl%
\url{https://github.com/iovisor/bpftrace}
\showURL{%
\tempurl}
\newblock
\shownote{Version 0.22.1; accessed February 04, 2025; licensed under
  Apache-2.0}.


\bibitem[Shan et~al\mbox{.}(2010)]%
        {a_programming_model_performance_study_shan}
\bibfield{author}{\bibinfo{person}{Hongzhang Shan}, \bibinfo{person}{Filip
  Blagojevi\'{c}}, \bibinfo{person}{Seung-Jai Min}, \bibinfo{person}{Paul
  Hargrove}, \bibinfo{person}{Haoqiang Jin}, \bibinfo{person}{Karl Fuerlinger},
  \bibinfo{person}{Alice Koniges}, {and} \bibinfo{person}{Nicholas~J. Wright}.}
  \bibinfo{year}{2010}\natexlab{}.
\newblock \showarticletitle{A programming model performance study using the NAS
  parallel benchmarks}.
\newblock \bibinfo{journal}{\emph{Sci. Program.}} \bibinfo{volume}{18},
  \bibinfo{number}{3–4} (\bibinfo{date}{Aug.} \bibinfo{year}{2010}),
  \bibinfo{pages}{153–167}.
\newblock
\urldef\tempurl%
\url{https://doi.org/10.1155/2010/715637}
\showDOI{\tempurl}


\bibitem[Shen et~al\mbox{.}(2024)]%
        {llamp_shen}
\bibfield{author}{\bibinfo{person}{Siyuan Shen}, \bibinfo{person}{Langwen
  Huang}, \bibinfo{person}{Marcin Chrapek}, \bibinfo{person}{Timo Schneider},
  \bibinfo{person}{Jai Dayal}, \bibinfo{person}{Manisha Gajbe},
  \bibinfo{person}{Robert Wisniewski}, {and} \bibinfo{person}{Torsten
  Hoefler}.} \bibinfo{year}{2024}\natexlab{}.
\newblock \showarticletitle{LLAMP: Assessing Network Latency Tolerance of HPC
  Applications with Linear Programming}. In \bibinfo{booktitle}{\emph{SC24:
  International Conference for High Performance Computing, Networking, Storage
  and Analysis}}. \bibinfo{pages}{1--18}.
\newblock
\urldef\tempurl%
\url{https://doi.org/10.1109/SC41406.2024.00070}
\showDOI{\tempurl}


\bibitem[Shende and Malony(2006)]%
        {tau_shende}
\bibfield{author}{\bibinfo{person}{Sameer~S. Shende} {and}
  \bibinfo{person}{Allen~D. Malony}.} \bibinfo{year}{2006}\natexlab{}.
\newblock \showarticletitle{The Tau Parallel Performance System}.
\newblock \bibinfo{journal}{\emph{The International Journal of High Performance
  Computing Applications}}  \bibinfo{volume}{20} (\bibinfo{date}{5}
  \bibinfo{year}{2006}), \bibinfo{pages}{287--311}.
\newblock
Issue 2.
\showISSN{1094-3420}
\urldef\tempurl%
\url{https://doi.org/10.1177/1094342006064482}
\showDOI{\tempurl}


\bibitem[Sridharan et~al\mbox{.}(2023)]%
        {chakra_sridharan}
\bibfield{author}{\bibinfo{person}{Srinivas Sridharan},
  \bibinfo{person}{Taekyung Heo}, \bibinfo{person}{Louis Feng},
  \bibinfo{person}{Zhaodong Wang}, \bibinfo{person}{Matt Bergeron},
  \bibinfo{person}{Wenyin Fu}, \bibinfo{person}{Shengbao Zheng},
  \bibinfo{person}{Brian Coutinho}, \bibinfo{person}{Saeed Rashidi},
  \bibinfo{person}{Changhai Man}, {and} \bibinfo{person}{Tushar Krishna}.}
  \bibinfo{year}{2023}\natexlab{}.
\newblock \bibinfo{title}{Chakra: Advancing Performance Benchmarking and
  Co-design using Standardized Execution Traces}.
\newblock
\newblock
\showeprint[arxiv]{2305.14516}~[cs.LG]
\urldef\tempurl%
\url{https://arxiv.org/abs/2305.14516}
\showURL{%
\tempurl}


\bibitem[sstsimulator(2025)]%
        {sst_dumpi_github}
\bibfield{author}{\bibinfo{person}{sstsimulator}.}
  \bibinfo{year}{2025}\natexlab{}.
\newblock \bibinfo{title}{{SST-DUMPI Trace Library}}.
\newblock
  \bibinfo{howpublished}{\url{https://github.com/sstsimulator/sst-dumpi}}.
\newblock
\newblock
\shownote{Accessed: 2025-02-14}.


\bibitem[Thompson et~al\mbox{.}(2022)]%
        {lammps_thompson}
\bibfield{author}{\bibinfo{person}{A.~P. Thompson}, \bibinfo{person}{H.~M.
  Aktulga}, \bibinfo{person}{R. Berger}, \bibinfo{person}{D.~S. Bolintineanu},
  \bibinfo{person}{W.~M. Brown}, \bibinfo{person}{P.~S. Crozier},
  \bibinfo{person}{P.~J. in~'t Veld}, \bibinfo{person}{A. Kohlmeyer},
  \bibinfo{person}{S.~G. Moore}, \bibinfo{person}{T.~D. Nguyen},
  \bibinfo{person}{R. Shan}, \bibinfo{person}{M.~J. Stevens},
  \bibinfo{person}{J. Tranchida}, \bibinfo{person}{C. Trott}, {and}
  \bibinfo{person}{S.~J. Plimpton}.} \bibinfo{year}{2022}\natexlab{}.
\newblock \showarticletitle{{LAMMPS} - a flexible simulation tool for
  particle-based materials modeling at the atomic, meso, and continuum scales}.
\newblock \bibinfo{journal}{\emph{Comp. Phys. Comm.}}  \bibinfo{volume}{271}
  (\bibinfo{year}{2022}), \bibinfo{pages}{108171}.
\newblock
\urldef\tempurl%
\url{https://doi.org/10.1016/j.cpc.2021.108171}
\showDOI{\tempurl}


\bibitem[Tikir et~al\mbox{.}(2009)]%
        {psins_tikir}
\bibfield{author}{\bibinfo{person}{Mustafa~M. Tikir},
  \bibinfo{person}{Michael~A. Laurenzano}, \bibinfo{person}{Laura Carrington},
  {and} \bibinfo{person}{Allan Snavely}.} \bibinfo{year}{2009}\natexlab{}.
\newblock \showarticletitle{PSINS: An Open Source Event Tracer and Execution
  Simulator}. In \bibinfo{booktitle}{\emph{2009 DoD High Performance Computing
  Modernization Program Users Group Conference}}. \bibinfo{pages}{444--449}.
\newblock
\urldef\tempurl%
\url{https://doi.org/10.1109/HPCMP-UGC.2009.73}
\showDOI{\tempurl}


\bibitem[Touvron et~al\mbox{.}(2023)]%
        {llama_touvron}
\bibfield{author}{\bibinfo{person}{Hugo Touvron}, \bibinfo{person}{Thibaut
  Lavril}, \bibinfo{person}{Gautier Izacard}, \bibinfo{person}{Xavier
  Martinet}, \bibinfo{person}{Marie-Anne Lachaux}, \bibinfo{person}{Timothée
  Lacroix}, \bibinfo{person}{Baptiste Rozière}, \bibinfo{person}{Naman Goyal},
  \bibinfo{person}{Eric Hambro}, \bibinfo{person}{Faisal Azhar},
  \bibinfo{person}{Aurelien Rodriguez}, \bibinfo{person}{Armand Joulin},
  \bibinfo{person}{Edouard Grave}, {and} \bibinfo{person}{Guillaume Lample}.}
  \bibinfo{year}{2023}\natexlab{}.
\newblock \bibinfo{title}{LLaMA: Open and Efficient Foundation Language
  Models}.
\newblock
\newblock
\showeprint[arxiv]{2302.13971}~[cs.CL]
\urldef\tempurl%
\url{https://arxiv.org/abs/2302.13971}
\showURL{%
\tempurl}


\bibitem[{University of Massachusetts Amherst}(2016)]%
        {umass_storage_trace}
\bibfield{author}{\bibinfo{person}{{University of Massachusetts Amherst}}.}
  \bibinfo{year}{2016}\natexlab{}.
\newblock \bibinfo{title}{UMass Trace Repository: Storage}.
\newblock
  \bibinfo{howpublished}{\url{https://traces.cs.umass.edu/docs/traces/storage/}}.
\newblock
\newblock
\shownote{Accessed: 4 February 2025. Copyright © 2016--2024 University of
  Massachusetts Amherst.}.


\bibitem[Vanini et~al\mbox{.}(2017)]%
        {flowlet_switching_vanini}
\bibfield{author}{\bibinfo{person}{Erico Vanini}, \bibinfo{person}{Rong Pan},
  \bibinfo{person}{Mohammad Alizadeh}, \bibinfo{person}{Parvin Taheri}, {and}
  \bibinfo{person}{Tom Edsall}.} \bibinfo{year}{2017}\natexlab{}.
\newblock \showarticletitle{Let It Flow: Resilient Asymmetric Load Balancing
  with Flowlet Switching}. In \bibinfo{booktitle}{\emph{14th USENIX Symposium
  on Networked Systems Design and Implementation (NSDI 17)}}.
  \bibinfo{publisher}{USENIX Association}, \bibinfo{address}{Boston, MA},
  \bibinfo{pages}{407--420}.
\newblock
\showISBNx{978-1-931971-37-9}
\urldef\tempurl%
\url{https://www.usenix.org/conference/nsdi17/technical-sessions/presentation/vanini}
\showURL{%
\tempurl}


\bibitem[Varga and Hornig(2008)]%
        {omnet_varga}
\bibfield{author}{\bibinfo{person}{Andr\'{a}s Varga} {and}
  \bibinfo{person}{Rudolf Hornig}.} \bibinfo{year}{2008}\natexlab{}.
\newblock \showarticletitle{An overview of the OMNeT++ simulation environment}.
  In \bibinfo{booktitle}{\emph{Proceedings of the 1st International Conference
  on Simulation Tools and Techniques for Communications, Networks and Systems
  \& Workshops}} (Marseille, France) \emph{(\bibinfo{series}{Simutools '08})}.
  \bibinfo{publisher}{ICST (Institute for Computer Sciences, Social-Informatics
  and Telecommunications Engineering)}, \bibinfo{address}{Brussels, BEL},
  Article \bibinfo{articleno}{60}, \bibinfo{numpages}{10}~pages.
\newblock
\showISBNx{9789639799202}


\bibitem[Wagner({[n.\,d.]})]%
        {meta_ai_center_wagner}
\bibfield{author}{\bibinfo{person}{Kurt Wagner}.}
  \bibinfo{year}{[n.\,d.]}\natexlab{}.
\newblock \showarticletitle{Meta Is Building New \$800 Million AI-Focused Data
  Center in Indiana}.
\newblock \bibinfo{journal}{\emph{Bloomberg}} (\bibinfo{year}{[n.\,d.]}).
\newblock
\urldef\tempurl%
\url{https://www.bloomberg.com/news/articles/2024-01-25/meta-building-new-800-million-ai-focused-data-center-in-indiana?embedded-checkout=true}
\showURL{%
\tempurl}


\bibitem[Wang et~al\mbox{.}(2025)]%
        {simai_wang}
\bibfield{author}{\bibinfo{person}{Xizheng Wang}, \bibinfo{person}{Qingxu Li},
  \bibinfo{person}{Yichi Xu}, \bibinfo{person}{Gang Lu}, \bibinfo{person}{Dan
  Li}, \bibinfo{person}{Chen Li}, \bibinfo{person}{Heyang Zhou},
  \bibinfo{person}{Linkang Zheng}, \bibinfo{person}{Sen Zhang},
  \bibinfo{person}{Yikai Zhu}, \bibinfo{person}{Yang Liu},
  \bibinfo{person}{Pengcheng Zhang}, \bibinfo{person}{Kun Qian}, {and}
  \bibinfo{person}{Kunling He}.} \bibinfo{year}{2025}\natexlab{}.
\newblock \showarticletitle{SimAI: Unifying Architecture Design and Performance
  Tunning for Large-Scale Large Language Model Training with Scalability and
  Precision}. In \bibinfo{booktitle}{\emph{22nd USENIX Symposium on Networked
  Systems Design and Implementation (NSDI 25)}}. \bibinfo{publisher}{USENIX
  Association}.
\newblock


\bibitem[Won et~al\mbox{.}(2023)]%
        {astra-sim2_won}
\bibfield{author}{\bibinfo{person}{William Won}, \bibinfo{person}{Taekyung
  Heo}, \bibinfo{person}{Saeed Rashidi}, \bibinfo{person}{Srinivas Sridharan},
  \bibinfo{person}{Sudarshan Srinivasan}, {and} \bibinfo{person}{Tushar
  Krishna}.} \bibinfo{year}{2023}\natexlab{}.
\newblock \bibinfo{title}{ASTRA-sim2.0: Modeling Hierarchical Networks and
  Disaggregated Systems for Large-model Training at Scale}.
\newblock
\newblock
\showeprint[arxiv]{2303.14006}~[cs.DC]
\urldef\tempurl%
\url{https://arxiv.org/abs/2303.14006}
\showURL{%
\tempurl}


\bibitem[Zahid et~al\mbox{.}(2017)]%
        {multi_tenant_hpc_zahid}
\bibfield{author}{\bibinfo{person}{Feroz Zahid}, \bibinfo{person}{Ernst~Gunnar
  Gran}, \bibinfo{person}{Bartosz Bogda{\'n}ski}, \bibinfo{person}{Bj{\o}rn~Dag
  Johnsen}, {and} \bibinfo{person}{Tor Skeie}.}
  \bibinfo{year}{2017}\natexlab{}.
\newblock \showarticletitle{Efficient network isolation and load balancing in
  multi-tenant HPC clusters}.
\newblock \bibinfo{journal}{\emph{Future Generation Computer Systems}}
  \bibinfo{volume}{72} (\bibinfo{year}{2017}), \bibinfo{pages}{145--162}.
\newblock
\showISSN{0167-739X}
\urldef\tempurl%
\url{https://doi.org/10.1016/j.future.2016.04.003}
\showDOI{\tempurl}


\bibitem[Zaitlen(2021)]%
        {nvtx_zaitlen}
\bibfield{author}{\bibinfo{person}{Ben Zaitlen}.}
  \bibinfo{year}{2021}\natexlab{}.
\newblock \bibinfo{title}{NVIDIA Tools Extension API: An Annotation Tool for
  Profiling Code in Python and C/C++}.
\newblock
\newblock
\urldef\tempurl%
\url{https://developer.nvidia.com/blog/nvidia-tools-extension-api-nvtx-annotation-tool-for-profiling-code-in-python-and-c-c/}
\showURL{%
\tempurl}
\newblock
\shownote{Accessed: 2025-01-28}.


\bibitem[Zhai et~al\mbox{.}(2010)]%
        {phantom_zhai}
\bibfield{author}{\bibinfo{person}{Jidong Zhai}, \bibinfo{person}{Wenguang
  Chen}, {and} \bibinfo{person}{Weimin Zheng}.}
  \bibinfo{year}{2010}\natexlab{}.
\newblock \showarticletitle{PHANTOM: predicting performance of parallel
  applications on large-scale parallel machines using a single node}. In
  \bibinfo{booktitle}{\emph{Proceedings of the 15th ACM SIGPLAN Symposium on
  Principles and Practice of Parallel Programming}} (Bangalore, India)
  \emph{(\bibinfo{series}{PPoPP '10})}. \bibinfo{publisher}{Association for
  Computing Machinery}, \bibinfo{address}{New York, NY, USA},
  \bibinfo{pages}{305–314}.
\newblock
\showISBNx{9781605588773}
\urldef\tempurl%
\url{https://doi.org/10.1145/1693453.1693493}
\showDOI{\tempurl}


\bibitem[Zhang et~al\mbox{.}(2022)]%
        {alibaba_zhang}
\bibfield{author}{\bibinfo{person}{Tianqi Zhang}, \bibinfo{person}{Yanqi
  Zhang}, \bibinfo{person}{Yuandong Tian}, \bibinfo{person}{Lin Ma},
  \bibinfo{person}{Wei Lin}, {and} \bibinfo{person}{Bin Cui}.}
  \bibinfo{year}{2022}\natexlab{}.
\newblock \showarticletitle{MLaaS in the Wild: Workload Analysis and Scheduling
  in Large-scale Heterogeneous GPU Clusters}. In \bibinfo{booktitle}{\emph{19th
  USENIX Symposium on Networked Systems Design and Implementation (NSDI 22)}}.
  \bibinfo{pages}{827--841}.
\newblock
\urldef\tempurl%
\url{https://github.com/alibaba/clusterdata/blob/master/cluster-trace-gpu-v2020/README.md}
\showURL{%
\tempurl}


\bibitem[Zheng et~al\mbox{.}(2004)]%
        {bigsim_zheng}
\bibfield{author}{\bibinfo{person}{G. Zheng}, \bibinfo{person}{Gunavardhan
  Kakulapati}, {and} \bibinfo{person}{L.V. Kale}.}
  \bibinfo{year}{2004}\natexlab{}.
\newblock \showarticletitle{BigSim: a parallel simulator for performance
  prediction of extremely large parallel machines}. In
  \bibinfo{booktitle}{\emph{18th International Parallel and Distributed
  Processing Symposium, 2004. Proceedings.}} \bibinfo{pages}{78--}.
\newblock
\urldef\tempurl%
\url{https://doi.org/10.1109/IPDPS.2004.1303013}
\showDOI{\tempurl}


\end{thebibliography}

\newpage
\appendix

\end{document}